\documentclass[11pt]{article}
 \usepackage[a4paper,hmarginratio=1:1,vmarginratio=2:3,
totalwidth=15.6cm,  totalheight=21.22cm]{geometry}
\renewcommand{\baselinestretch}{1.37}
\usepackage{amsmath}
\usepackage{amssymb}
\usepackage{amsfonts}
\usepackage{epsfig}
\usepackage{graphics}
\newcommand{\si}{{\sigma}}

\newcommand{\cL}{{\cal L}}

\newcommand{\cO}{{\cal O}}

\newcommand{\be}{\begin{equation}}
\newcommand{\ee}{\end{equation}}
\newcommand{\bea}{\begin{eqnarray}}
\newcommand{\eea}{\end{eqnarray}}

\newcounter{oldcounter}
\addtocounter{equation}{1}
\begin{document} 

\begin{flushright} 
{OUTP-09-07P}
\end{flushright}

\thispagestyle{empty} \vspace{3cm} 
 
\begin{center} 
{\Large \textbf{Fine tuning as an indication of physics beyond the MSSM}} 
\bigskip 
 
\vspace{1.cm} \textbf{
S. Cassel \footnote{\,e-mail: s.cassel1@physics.ox.ac.uk}, 
D.~M. Ghilencea \footnote{\,e-mail: d.ghilencea1@physics.ox.ac.uk}, 
G. G. Ross \footnote{\,e-mail: g.ross1@physics.ox.ac.uk}}\\[0pt] 
\vspace{0.5cm} { Rudolf\, Peierls\, Centre for Theoretical Physics, 
\,University\, of\, Oxford,\\[0pt] 
1 Keble Road, Oxford OX1 3NP, United Kingdom.}\\[6pt] 
\end{center} 
 
\medskip 
\begin{abstract}

\noindent We investigate the amount of fine tuning of the electroweak scale 
in the presence of new physics beyond the MSSM, parametrized by 
higher dimensional operators. We show that these significantly reduce the 
MSSM fine tuning to $\Delta<10$ 
for a Higgs mass between the LEPII bound and $130$ GeV, and 
a corresponding scale $M_{\ast }$ of new physics as high as 
$30$ to $65$ times the Higgsino mass. If the fine-tuning criterion
 is indeed of physical relevance, 
the findings indicate the presence of new physics in the form of new states 
of mass of $\cO(M_{\ast })$ that generated the effective operators in the 
first instance. At small $\tan \beta $ these states can be a gauge singlet
or a $SU(2)$ triplet. 
We derive analytical results for the EW scale fine-tuning for the MSSM with 
higher dimensional operators, including the quantum corrections which are 
also applicable to the pure MSSM case in the limit the coefficients of the 
higher dimension operators vanish. A general expression for the fine-tuning 
is also obtained for an arbitrary two-Higgs doublet potential. 
\end{abstract}

\newpage 
\setcounter{page}{1} 
\def\thefootnote{\arabic{footnote}}
\setcounter{footnote}{0}

\section{Introduction.} 
 
Low-energy supersymmetry offers an elegant solution to the hierarchy 
problem. One consequence of this is that it introduces a spectrum of 
supersymmetric states in the visible sector with mass of the order of the 
electroweak scale. However, none of the superpartners of the Standard Model 
have been seen, although there is hope that LHC will soon remedy this.  
In trying to determine the physics beyond the Standard Model the 
fact that no superpartners have been observed is significant as it 
(re)introduces the need for some amount of fine tuning of the parameters of 
the minimal supersymmetric standard model (MSSM), to separate the 
electroweak and supersymmetry breaking scales (the \textquotedblleft little 
hierarchy problem\textquotedblright ). On the other hand 
circumstantial evidence for supersymmetry such as the successful unification 
of couplings
 \cite{Dimopoulos:1981yj,Dimopoulos:1981zb,Ibanez:1981yh,Ghilencea:2001qq} or 
radiative electroweak breaking \cite{Ibanez:1982fr} is consistent with and 
in fact requires the existence of such light superpartners.

The basic issue raised by fine-tuning is the sensitivity of the electroweak 
scale (more precisely the mass of Z) to small variations of the input 
parameters of the MSSM, consistent with the measured Z mass, $m_{Z},$ and 
the current bounds on the lightest Higgs mass, $m_{h}$. The need to 
keep fine tuning small indicates a light Higgs in some tension with the 
LEPII bound \cite{higgsboundLEP} $m_{h}\geq 114.4$  GeV.  
Consistency of this bound with the MSSM tree level bound $m_{h}\leq m_{Z}$ 
can only be achieved at the quantum level, by a large top quark/squarks loop 
correction to $m_{h}$. To maximise this, the top squarks must be quite 
massive or highly mixed implying that the MSSM model is fine-tuned.

Various definitions of fine tuning have been proposed. The most 
popular one \cite{r3} is based on the logarithmic derivatives of the 
observables with respect to the set of parameters considered. It has been 
widely used in quantifying the fine tuning in the MSSM 
\cite{Barbieri:1998uv,Chankowski:1998xv,Chankowski:1997zh,Kane:1998im, 
Batra:2003nj,Giudice:2006sn,Casas:2003jx}
and we shall use it in this analysis. The new 
feature of our analysis starts with the premise that if low-energy 
supersymmetry is indeed the solution to the hierarchy problem, a significant 
amount of fine tuning of the electroweak scale in the MSSM may in 
fact suggest that there are additional new degrees of freedom in the theory 
beyond those of the MSSM. There are many models that consider additional 
degrees of freedom beyond the MSSM in order to reduce the amount of fine 
tuning. The NMSSM is just such an example \cite{Dermisek:2007yt} which has 
an extra chiral singlet. One can consider other MSSM extensions with more 
chiral fields, additional gauge interactions, etc. Each of these brings 
different solutions to the little hierarchy problem and it is difficult to 
assess which of these is the most compelling. In this paper we 
perform a model independent analysis of the nature of this new physics based 
on a general parametrisation of physics beyond the MSSM. In particular we 
extend the MSSM by the addition of higher dimensional operators 
\cite{Piriz:1997id,Polonsky:2000zt,M0,Barbieri:1999tm}
 that encode the effect of 
all possible new physics at scales below the appearance of the new degrees 
of freedom. Having identified the most relevant operators one can later 
address the question of what new physics generated these operators in the 
first instance. The advantage of the effective approach  is that it 
provides an organising principle according to which one usually restricts 
the analysis to operators of a given (leading) order in the scale of new 
physics $M_{\ast }$, with higher order operators suppressed by higher powers 
of $M_{\ast }$. The analysis we consider includes dimension $d=5$ and $d=6$
operators beyond MSSM \cite{Dine,Antoniadis:2008es,Blum:2008ym}. For 
the case of $d=5$  operators we determine the amount of fine tuning 
as a function of the mass of the lightest Higgs corrected by the quantum 
contributions using both analytical and numerical techniques. One-loop 
renormalisation group corrections in the Higgs potential are also included.

The expectation that higher dimensional operators can reduce the amount of 
fine tuning is broadly based on two arguments. Firstly these operators may 
directly increase the tree level value of $m_{h}$ \cite%
{Dine,Antoniadis:2008es}. Consequently the tree level upper bound on $m_{h}$ 
may be relaxed, and the quantum effects needed to satisfy LEPII bound may be 
smaller, corresponding to reduced fine tuning.  Secondly, the higher 
dimensional operators may generate additional contributions to the quartic 
Higgs couplings of the MSSM, again serving to increase the Higgs mass. This 
effect can be quite significant because, in the MSSM, the quartic coupling,  
$(g_{2}^{2}+g_{1}^{2})/8,$ where $g_{2}$ and $g_{1}$ are the $SU(2)\times 
U(1)\ $gauge couplings, is anomalously small; indeed its smallness 
is a major source of the little hierarchy problem. For the case of just the 
$d=5$ operators a numerical study shows that these effects can 
reduce the amount of fine tuning, $\Delta ,$  of the electroweak 
(EW) scale relative to the MSSM case, to less than $10$ for a 
Higgs mass in the range  $114.4\,$GeV\,$\leq m_{h} \leq 130$ GeV
even for a scale of new physics as high as $(30\,\, \mathrm{to}\,\, 65)$
 times the higgsino mass, and possibly above the LHC reach. We also give
 in Appendix an analytical formula for 
the EW fine-tuning in a general two-Higgs doublet model, which can be easily 
applied to specific models.

The plan of the paper is as follows. Section~\ref{s1} lists the $d=5$ and $%
d=6$ operators that are consistent with the MSSM symmetries and that can 
affect fine tuning. In Section~\ref{finetuning} we evaluate analytically and 
numerically the fine tuning in the MSSM extended by the  $d=5$ operators.\  
The conclusions are given in Section~\ref{conclusions}.

\section{Higher dimensional operators beyond MSSM Higgs sector} 
 
\label{s1} 
 
In this section we list the effective operators of dimension $d=5,6$ that 
can be present in the Higgs sector consistent with the symmetries of the 
MSSM. These operators parametrise new physics beyond the MSSM and affect the 
Higgs scalar potential. Therefore they also affect the amount of fine tuning 
of the EW scale, as discussed in detail in the next section. The 
($R-$parity conserving) $d=5$ operators in the MSSM Higgs sector are: 
\medskip  
\begin{eqnarray} 
\mathcal{L}_{1} &=&\frac{1}{M_{\ast }}\int d^{2}\theta \,\lambda 
(S)\,(H_{1}\,H_{2})^{2},  \label{dimensionfive} \\[0.06in] 
\mathcal{L}_{2} &=&\frac{1}{M_{\ast }}\int d^{4}\theta \,\,\Big\{%
\,A(S,S^{\dagger })D^{\alpha }\Big[B(S,S^{\dagger })\,H_{2}\,e^{-V_{1}}\Big]%
D_{\alpha }\Big[C(S,S^{\dagger })\,e^{V_{1}}\,H_{1}\Big] 
+h.c.\Big\} 
\end{eqnarray}

\medskip\noindent
where $S$ is the spurion field, $S=\theta \theta \,m_{0}$,
$A(S,S^{\dagger })$, $B(S,S^{\dagger })$, $C(S,S^{\dagger })$ are 
polynomials in $S,S^{\dagger }$ 
 and $m_{0}$ is the susy breaking scale in the visible 
sector (in gravity mediation $m_{0}=\langle F_{h}\rangle /M_{Planck}$
where 
$\langle F_{h}\rangle $ is the auxiliary field vacuum expectation value (vev) 
in the hidden sector responsible for supersymmetry breaking). As we discuss 
in Section \ref{origin} the first operator can be generated, for example, by 
integrating out massive gauge singlets or $SU(2)$ triplets, while the second 
is easily generated by integrating out a pair of massive Higgs
doublets  \cite{Antoniadis:2008es}, all of mass of order $M_{\ast }$.

In \cite{Antoniadis:2008es,Antoniadis:2009rn}
 it was shown that by using general 
field redefinitions one can remove $\mathcal{L}_{2}$ from the action. The 
effect of this is an overall renormalisation of the soft terms and of the $%
\mu $ term. Since the fine tuning measure includes the fine tuning 
with respect to each of these soft operators separately adding $\cL_{2}$
cannot reduce the overall fine tuning. For this reason we will 
only include $\cL_{1}$  in our discussion of fine tuning with $d=5$
operators. 
 
There are also $d=6$ operators that can be present in addition to the MSSM 
Higgs sector. These are suppressed relative to the $d=5$ operators by the 
factor $1/M_{\ast }$. However they may give contributions to the 
Higgs potential enhanced by $\tan \beta $  relative to the $d=5$
 so cannot be ignored at very large $\tan \beta $. The 
list of $d=6$ operators  is  (see also \cite{Piriz:1997id,Polonsky:2000zt}): 
\medskip  
\begin{eqnarray} 
\mathcal{O}_{i} &=&\frac{1}{M_{\ast }^{2}}\int d^{4}\theta \,\,\mathcal{Z}%
_{i}(S,S^{\dagger })\,\,(H_{i}^{\dagger }\,e^{V_{i}}\,H_{i})^{2},\qquad 
i=1,2.\qquad   \nonumber \\ 
\mathcal{O}_{3} &=&\frac{1}{M_{\ast }^{2}}\int d^{4}\theta \,\,\mathcal{Z}%
_{3}(S,S^{\dagger })\,\,(H_{1}^{\dagger 
}\,e^{V_{1}}\,H_{1})\,(H_{2}^{\dagger }\,e^{V_{2}}\,H_{2}),\qquad  
\label{dim6} 
\end{eqnarray}

\medskip\noindent
(These can be generated by integrating a massive $U(1)$ gauge boson or
a $SU(2)$ triplet).
\medskip
\begin{eqnarray} 
\mathcal{O}_{4} &=&\frac{1}{M_{\ast }^{2}}\int d^{4}\theta \,\,\mathcal{Z}%
_{4}(S,S^{\dagger })\,\,(H_{2}\,H_{1})\,(H_{2}\,H_{1})^{\dagger },\quad 
\qquad \qquad   \nonumber \\
\mathcal{O}_{5} &=&\frac{1}{M_{\ast }^{2}}\int d^{4}\theta \,\,\mathcal{Z}%
_{5}(S,S^{\dagger })\,\,(H_{1}^{\dagger 
}\,e^{V_{1}}\,H_{1})\,(H_{2}\,H_{1}+h.c.)  \nonumber \\
\mathcal{O}_{6} &=&\frac{1}{M_{\ast }^{2}}\int d^{4}\theta \,\,\mathcal{Z}%
_{6}(S,S^{\dagger })\,\,(H_{2}^{\dagger 
}\,e^{V_{2}}\,H_{2})\,(H_{2}\,H_{1}+h.c.)  \nonumber \\ 
\mathcal{O}_{7} &=&\frac{1}{M_{\ast }^{2}}\int d^{2}\theta \,\,\mathcal{Z}%
_{7}(S,0)\,\,W^{\alpha }\,W_{\alpha }\,(H_{2}\,H_{1})+h.c.,  \nonumber 
\\
\mathcal{O}_{8} &=&\frac{1}{M_{\ast }^{2}}\int d^{4}\theta
 \,\,\Big[\mathcal{Z}_{8}(0,S^{\dagger })\,\,(H_{2}\,H_{1})^{2}+h.c.\Big] 
\end{eqnarray}%

\medskip\noindent
where $W_{\alpha }$ is the supersymmetric field strength of a vector 
superfield of the SM gauge group. $\cO_4$ can for example be generated
 by integrating a gauge singlet. 
\medskip  
\begin{eqnarray} 
\mathcal{O}_{9} &=&\frac{1}{M_{\ast }^{2}}\int d^{4}\theta \,\,\mathcal{Z} 
_{9}(S,S^{\dagger })\,\,H_{1}^{\dagger }\,\overline{\nabla }
^{2}\,e^{V_{1}}\,\nabla ^{2}\,H_{1}  \nonumber \\ 
\mathcal{O}_{10} &=&\frac{1}{M_{\ast }^{2}}\int d^{4}\theta \,\,\mathcal{Z}
_{10}(S,S^{\dagger })\,\,H_{2}^{\dagger }\,\overline{\nabla } 
^{2}\,e^{V_{2}}\,\nabla ^{2}\,H_{2}  \nonumber \\ 
\mathcal{O}_{11} &=&\frac{1}{M_{\ast }^{2}}\int d^{4}\theta \,\,\mathcal{Z}
_{11}(S,S^{\dagger })\,\,H_{1}^{\dagger }\,e^{V_{1}}\,\nabla ^{\alpha 
}\,W_{\alpha }\,H_{1}  \nonumber \\ 
\mathcal{O}_{12} &=&\frac{1}{M_{\ast }^{2}}\int d^{4}\theta \,\,\mathcal{Z} 
_{12}(S,S^{\dagger })\,\,H_{2}^{\dagger }\,e^{V_{2}}\,\nabla ^{\alpha 
}\,W_{\alpha }\,H_{2}  \label{der} 
\end{eqnarray}

\medskip\noindent
where $\nabla _{\alpha }$ acts on everything to the right and $\nabla 
_{\alpha }\,H_{i}=e^{-V_{i}}\,D_{\alpha }\,e^{V_{i}} H_i$. i=1,2. In addition to 
the spurion dependence in the wavefunctions $\mathcal{Z}_{i}(S,S^{\dagger })$ 
, extra $(S,S^{\dagger })$ dependence (not shown) can be present under each 
derivative $\nabla _{\alpha }$ in eq.(\ref{der}), in order to ensure the most 
general supersymmetry breaking contribution associated to these operators. 
One may use the equations of motion to replace the operators 
involving  extra derivatives by non-derivative ones\footnote{
Setting higher derivative operators onshell is a subtle issue in this case.
One can also use general 
spurion-dependent field redefinitions to ``gauge 
away'' (some of) these operators,  using the method
 of~\cite{Antoniadis:2008es,Antoniadis:2009rn}.}.
 Note that when computing the fine tuning 
measure eliminating a particular operator will lead to correlations 
between the remaining operators that, strictly, should be taken into 
account. 
 
Given the large number of $d=6$ operators, determination 
of the fine tuning with respect to their coefficients is difficult. For this 
reason we restrict the following discussion to the  $d=5$
operators. In Section~\ref{largetanbeta} we comment on the new contributions 
that may come from $d=6$ operators and discuss the limit on our 
analysis that follows from keeping only $d=5$  operators.

\section{Fine-tuning in MSSM with d=5 operators (MSSM$_5$).} 
 \label{finetuning} 
 
\subsection{The scalar potential} 
\label{general} 
 
In this section we evaluate the EW scale
fine-tuning in the MSSM extended by  
$\mathcal{L}_{1}$ in eq.(\ref{dimensionfive}).
Including it together with the 
MSSM, the full Higgs Lagrangian is then given by  
\medskip
\begin{eqnarray} 
\mathcal{L} &=&\int d^{4}\theta \,\Big[\,(1-c_{h_{1}}\,S\,S^{\dagger 
})\,H_{1}^{\dagger }\,e^{V_{1}}\,H_{1}+(1-c_{h_{2}}\,S\,S^{\dagger 
})\,H_{2}^{\dagger }\,e^{V_{2}}\,H_{2}\Big]  \nonumber \\[4pt] 
&+&\int d^{2}\theta \,\Big[\mu _{0}\,(1+B_{0}\,S)\,H_{1}\,H_{2}+
\frac{1}{M_{\ast }}\,(1+c_{0}S)\,(H_{1}\,H_{2})^{2}\Big]+h.c. 
\end{eqnarray} 

\noindent\medskip
The corresponding scalar potential is given by
\medskip  
\begin{eqnarray} 
V &=&\tilde{m}_{1}^{2}\,|h_{1}\,|^{2}+\tilde{m}_{2}^{2}\,|h_{2}\,|^{2}-\big( 
m_{3}^{2}\,h_{1}\,h_{2}+h.c.\big)+\frac{g^{2}}{8}\,\big(|\,h_{1}\,|^{2}-|
\,h_{2}\,|^{2}\big)^{2}+\frac{g^{2}}{8}\,\delta \,|\,h_{2}\,|^{4} 
\label{scalarV} \\[5pt] 
&+&\big(|\,h_{1}\,|^{2}+|\,h_{2}\,|^{2}\big)\,\big(\zeta 
_{1}\,h_{1}\,h_{2}+h.c.\big)+\frac{1}{2}\,\big(\,\zeta 
_{2}\,(h_{1}\,h_{2})^{2}+h.c.\big)  \nonumber 
\end{eqnarray} 

\medskip\noindent
where $g^{2}\equiv g_{1}^{2}+g_{2}^{2}$,
\medskip
\begin{equation} 
\zeta _{1}=2\,\mu _{0}^{\ast }/M_{\ast },\qquad
\zeta _{2}=-2\,c_{0}\,m_{0}/M_{\ast 
}  \label{coeffs} 
\end{equation}
and
\begin{eqnarray} 
\tilde{m}_{1}^{2}(t) &=&m_{0}^{2}+\mu _{0}^{2}\,{\sigma } 
_{8}^{2}(t)+m_{12}^{2}\,\sigma _{1}(t)  \nonumber  \label{sm} \\[5pt] 
\tilde{m}_{2}^{2}(t) &=&\mu _{0}^{2}\,{\sigma
}_{8}^{2}(t)+m_{12}^{2}\,
{\sigma }_{4}(t)+A_{t}\,m_{0}\,m_{12}\,{\sigma }_{5}(t)+m_{0}^{2}\,{\sigma } 
_{7}(t)-m_{0}^{2}\,A_{t}^{2}\,{\sigma }_{6}(t)  \nonumber \\[5pt] 
m_{3}^{2}(t) &=&\mu _{0}\,m_{12}\,\sigma _{2}(t)+B_{0}\,m_{0}\,\mu _{0}\,{ 
\sigma }_{8}(t)+\mu _{0}\,m_{0}\,A_{t}\,\sigma _{3}(t) 
\end{eqnarray} 
 
\medskip \noindent 
The coefficients $\sigma _{i}$ depend on $t\equiv \ln 
M_{G}^{2}/Q^{2}$ with functional dependence given in 
\cite{r3,deBoer,r1,r2,r4}.
 The (high scale) boundary values ($t=0$) are normally 
chosen to be $\sigma _{1,2,..,6}=0$, $\sigma _{7,8}=1$ (i.e. $c_{h_{1,2}}=1$ 
). For $Q^{2}=m_{Z}^{2}$ ($t=t_{z}$) the values of these coefficients are 
given in Appendix~\ref{appendixA} in terms of the top Yukawa coupling. 
To simplify notation we will not display the argument $t_{z}$ in what 
follows.

The quartic term $\delta |h_{2}|^{4}$ is generated radiatively 
\cite{Giudice:2006sn,Carena:1995bx}. Including leading log two-loop
 effects one has 
\medskip  
\begin{eqnarray} 
\delta  &=&\frac{3\,h_{t}^{4}}{g^{2}\,\pi ^{2}\,}\bigg[\ln \frac{M_{\tilde{t}
}}{m_{t}}+\frac{X_{t}}{4}+\frac{1}{32\pi ^{2}}\,\Big(3\,h_{t}^{2}-16
\,g_{3}^{2}\Big)\Big(X_{t}+2\ln \frac{M_{\tilde{t}}}{m_{t}}\Big)\ln \frac{M_{
\tilde{t}}}{m_{t}}\bigg],  \nonumber
\eea
\bea
X_{t} &\equiv &\frac{2\,(A_{t}\,m_{0}-\mu \cot \beta )^{2}}{M_{\tilde{t}}^{2}
}\,\,\Big(1-\frac{(A_{t}\,m_{0}-\mu \cot \beta )^{2}}{12\,\,M_{\tilde{t}}^{2} 
}\,\Big). 
\end{eqnarray}

\medskip\noindent 
with $M_{\tilde{t}}^{2}\equiv m_{\tilde{t}_{1}}\,m_{\tilde{t}_{2}}$, 
and $g_{3}$  is the strong coupling. 

The minimum conditions for $V$ can be written as: 
\medskip  
\begin{equation} 
v^{2}=-\frac{m^{2}}{\lambda },\qquad \left( 2\lambda \frac{\partial m^{2}}{%
\partial \beta }-m^{2}\,\frac{\partial \lambda }{\partial \beta }\right) 
_{\beta =\beta _{\min }}=0,  \label{mincond} 
\end{equation} 
 
\medskip\noindent 
with the notation $v^{2}=v_{1}^{2}+v_{2}^{2}$, $\tan 
\beta =v_{2}/v_{1}$, $m_{Z}^{2}=g^{2}\,v^{2}/4$ and where: \medskip  
\begin{eqnarray} 
m^{2} &\equiv &\tilde{m}_{1}^{2}\,\cos ^{2}\beta +\tilde{m}_{2}^{2}\,\sin 
^{2}\beta -m_{3}^{2}\,\sin {2\beta }  \nonumber  \label{lambda} \\[8pt] 
\lambda  &\equiv &\frac{g^{2}}{8}\,(\cos ^{2}{2\beta }+\delta \,\sin 
^{4}\beta )+\zeta _{1}\,\sin {2\beta }+\frac{\zeta _{2}}{4}\,\sin ^{2}{%
2\beta .} 
\end{eqnarray} 
 
\medskip 
\noindent  
Note that in deriving these expressions we have discarded
non-leading log corrections except those to the quartic Higgs coupling 
where the tree level term is anomalously small.

\subsection{Analytical results for fine-tuning} 
 
The fine tuning of the EW scale with respect to a set of parameters $p$ 
introduced in \cite{r3} is 
\medskip  
\begin{equation} 
\Delta \equiv \max {\text{Abs}}\big[\Delta _{p}\big]\Big\vert_{p=\{\mu 
_{0}^{2},m_{0}^{2},A_{t}^{2},B_{0}^{2},m_{12}^{2},\,\zeta _{1}^{2},\,\zeta 
_{2}^{2}\}},\qquad \Delta _{p}\equiv \frac{\partial \ln
  v^{2}}{\partial \ln p}  \label{ft} 
\end{equation}

\medskip
\noindent
With $m^{2}=m^{2}(p,\beta )$, $\lambda =\lambda (p,\beta )$
we can find ${\partial \beta }/{\partial p}$ from 
the second of eqs.(\ref{mincond}) (more precisely this determines the 
parameter dependence of $\beta _{\min })$:
\medskip
\[ 
\frac{\partial \beta }{\partial p}=\frac{1}{z}\left( -2\frac{\partial 
\lambda }{\partial p}\frac{\partial m^{2}}{\partial \beta }-2\lambda \frac{ 
\partial ^{2}m^{2}}{\partial \beta \partial p}+\frac{\partial m^{2}}{
\partial p}\frac{\partial \lambda }{\partial \beta }+m^{2}\frac{\partial 
^{2}\lambda }{\partial \beta \partial p}\right)  
\]
\medskip\noindent
where
\medskip
\[ 
z=\lambda \,\bigg(2\frac{\partial ^{2}m^{2}}
{\partial \beta ^{2}}+v^{2}\frac{ 
\partial ^{2}\lambda }{\partial \beta ^{2}}\bigg)-\frac{v^{2}}{2}\,\bigg(
\frac{\partial \lambda }{\partial \beta }\bigg)^{2}. 
\]

\medskip\noindent
Using this one finds \cite{Casas:2003jx} 
 
\begin{equation} 
\Delta _{p}=-\frac{p}{z}\,\bigg[\bigg(2\frac{\partial ^{2}m^{2}}{\partial 
\beta ^{2}}+v^{2}\frac{\partial ^{2}\lambda }{\partial \beta ^{2}}\bigg)
\bigg(\frac{\partial \lambda }{\partial p}+\frac{1}{v^{2}}\frac{\partial 
m^{2}}{\partial p}\bigg)+\frac{\partial m^{2}}{\partial \beta }\frac{
\partial ^{2}\lambda }{\partial \beta \partial p}-\frac{\partial \lambda }{
\partial \beta }\frac{\partial ^{2}m^{2}}{\partial \beta \partial p}\bigg]. 
\label{delta2} 
\end{equation}

\subsubsection{A general two-Higgs model} 
 
Using eq.(\ref{delta2}) we derived a general analytical result 
for the fine-tuning of the EW scale in a general two-Higgs doublet 
model allowing for the most general renormalisable Higgs potential, 
see Appendix~\ref{appendixB}, eq.(\ref{ggg}). The results are
presented in terms of derivatives of the soft masses and couplings
of the scalar potential.

\bigskip

\subsubsection{The MSSM with dimension-five operators (MSSM$_5$)} 
 
Applied to the case of the MSSM with dimension-five operators  
the results   presented in  Appendix~\ref{appendixB}  give:

\bea
\label{mu}
\Delta_{\mu_0^2}
&=& -\frac{1}{v^2\,D}\,
\Big\{ v^2\,\cos 2\beta\,\Big[
\sin 2\beta\,
\Big(\zeta_1\, (\,2 \,\gamma_2 -\delta\, g^2\,v^2/8)
\nonumber\\[5pt]
&+& 2\, \gamma_1\, \big[\,\delta g^2/8
- \big( (4+\delta)\,g^2/8-\zeta_2\big)
\,\,\cos 2\beta\,\,\big]\Big)\Big]
+2 \Big[\,2\mu_0^2\,\si_8^2+(\zeta_1\,v^2-\gamma_1)
\sin 2\beta\Big]
\nonumber\\[5pt]
&\times &\!\!\!\!\! \Big[
\gamma_4 - v^2\,\big(2\zeta_1\sin2\beta- \zeta_2\cos4\beta\big)
\Big]\!\Big\}\,\,\,\,\,\,
\eea

\bea
\label{m0}
\Delta_{m_0^2}&=&
-\frac{1}{4\,v^2\,D}
\Big\{-v^2\zeta_2\sin 4\beta\, \Big[
4\gamma_1\,\,\cos2\beta+\big(\delta\,g^2\,v^2/8-2\gamma_2\big)
\sin2\beta\Big]
\nonumber\\[5pt]
&+&\!\!\!\! 2\,v^2\Big[2\,(\gamma_1\!-\mu_0\,m_{12}\sigma_2)
\cos 2\beta\!+\gamma_3\sin 2\beta\Big]
\Big[4\zeta_1\cos2\beta
+\!\delta g^2\cos\beta\sin^3\!\beta
\nonumber\\[5pt]
&+& (\zeta_2-g^2/2)\,\sin4\beta\Big]
+8\,\Big[
\gamma_4
- v^2\,\big(
2\zeta_1\sin2\beta-\zeta_2\cos4\beta
\big)\Big]
\nonumber\\[5pt]
&\times&\Big[2 m_0^2 -\gamma_3\,\sin^2\beta+ \zeta_2 \, 
v^2 \sin^2\beta\cos^2\beta
-(\gamma_1-m_{12}\,\mu_0\,\sigma_2
) \sin2\beta\Big]\Big\}\qquad\qquad\qquad
\eea

\bea
\label{m12}
\Delta_{m_{12}^2}&=&
-\frac{m_{12}}{v^2\,D}\,\,
\Big\{\,\frac{v^2}{2}\, \Big[ \, 2\mu_0\sigma_2\cos2\beta-
\big(A_t\,\si_5\,\,m_0+2\,m_{12}\,(\si_4-\sigma_1)\big)\,\sin 2\beta\Big]
\nonumber\\[5pt]
&\times&
\Big[4\zeta_1\cos2\beta  
+\delta \,g^2\cos\beta\sin^3\beta
+(\zeta_2-g^2/2)\,\sin4\beta\Big]
+ 2\,
\Big[2\,m_{12}\,\sigma_1-\mu_0\,\sigma_2\sin 2\beta
\nonumber\\[5pt]
&+&\big(A_t\,\si_5\,m_0+2\,m_{12}(\si_4-\sigma_1)\big)\sin^2\beta\Big]
\Big[\gamma_4
- v^2\,\big(
2\zeta_1\sin2\beta-\zeta_2\cos4\beta
\big)\Big]\Big\}\,\,\qquad\qquad\quad
\eea

\bea
\label{At}
\,\,\,\Delta_{A_t^2}&=&-\frac{A_t}{v^2\,D}\,\Big\{2\,m_0\sin
\beta\,\Big[2\mu_0\,\sigma_3\,\cos\beta
+(2\,A_t\,\si_6\,m_0-\si_5\,m_{12})\sin\beta\Big]
\Big[
-\zeta_2 \,v^2\, \cos 4\beta
\nonumber\\[5pt]
&+& 2\,\zeta_1 \,v^2\,\sin 2\beta-\gamma_4
\Big]
+m_0\,v^2\,\Big[\mu_0\sigma_3\cos
2\beta-(1/2) \si_5\,m_{12}\,\sin2\beta
+A_t\,\si_6\,m_0\sin 2\beta\Big]
\nonumber\\[5pt]
&\times& 
\Big[\,4\,\zeta_1\cos2\beta 
+\delta \,g^2\cos\beta\sin^3\beta
+(\zeta_2-g^2/2)\,\sin4\beta\Big]\Big\}
\eea

\bea
\Delta_{B_0^2}&=&
-\frac{2 \,B_0\,m_0\,\mu_0\, \si_8}{v^2\,D}\,
\Big\{v^2\,\Big[ 2\zeta_1+(\zeta_2-(4+\delta) g^2/8)
\sin^3 2\beta\Big]
\nonumber\\[5pt]
&&\qquad\qquad\qquad\qquad\qquad +\,\,
(\delta\, g^2\,v^2/16-\gamma_2)\,\sin4\beta
- 4\gamma_1\,\sin^2 2\beta \Big\}
\qquad\qquad\qquad
\eea
Also:
\bea
\Delta_{\zeta_1^2}&=&
 -\frac{\zeta_1}{2\,D'}\,
\Big[2\,m_3^2-2\cos 4\beta \,\,\big(3
  m_3^2+(4+\delta)\,(g^2 v^2/8) \,\sin 2\beta\big)
 \nonumber\\[5pt]
 &&\qquad\qquad\qquad\qquad\qquad +
\,\big(\, 3\, (\tilde m_2^2- \tilde m_1^2)+\delta g^2 v^2/8\big)\,
\sin 4\beta\Big]\qquad\qquad
\eea
\bea
\Delta_{\zeta_2^2}
&=& -\frac{\zeta_2}{8\,D'}
\sin 2\beta\, \Big[
-2\cos 4\beta\,\big(4 \,m_3^2+(4+\delta) \,\,(g^2 v^2/8)\,
\sin 2\beta\big)\qquad\qquad
\nonumber\\[5pt]
&&\qquad\qquad\qquad\qquad\quad\, +\,
\,\big( \,4\,\,(\tilde m_2^2-\tilde m_1^2)+\delta  g^2 v^2/8 \big)\,
\sin 4\beta\Big]
\eea

\medskip\noindent
with the notation:

\medskip
\bea
D&\equiv & 2\,\Big\{
-\frac{1}{8}\,v^2
\Big[
4\zeta_1\cos2\beta+\zeta_2\sin 4\beta
+g^2\,(\delta \cos\beta\sin^3\beta-1/2\,\sin4\beta)
\Big]^2\\[5pt]
&\!\!\!-&
\!\!\!\!\!
2\Big[\zeta_1\sin2\beta+\zeta_2/4\sin^2 2\beta+g^2/8\,
(\cos^2 2\beta+\delta\sin^4\beta)\Big]
\Big[
v^2\,\big(2\zeta_1\sin 2\beta-\zeta_2\cos
4\beta\big)
-\gamma_4 \Big]\Big\}\nonumber
\eea
\medskip
\bea
D'&\equiv &
\frac{g^2}{4}\,\big(\cos^2 2\beta+\delta \sin^4\beta\big)
\Big[
\big(\,2 \,(\tilde m_2^2-\tilde m_1^2)+\delta g^2 v^2/8\big)\,
\cos 2\beta
\nonumber\\[5pt]
&-&(4+\delta)\,(g^2 v^2/8)\,\cos 4\beta+4\, m_3^2 \sin
2\beta
\Big]
-\frac{g^4 v^2}{32}\,
\big[ 2-(4+\delta)\sin^2\beta\big]^2\,\sin^2 2\beta\qquad\quad\qquad
\eea

\bigskip\noindent
and
\begin{eqnarray} \label{gammaall}
\gamma _{1} &\equiv &\mu _{0}\,(B_{0}\,m_{0}\,{\sigma }_{8}+m_{12}\,\sigma 
_{2}+A_{t}\,m_{0}\,\sigma _{3})  \nonumber  
\label{gammas} \\
\gamma _{2} &\equiv &(-1+{\sigma }_{7}-A_{t}^{2}\,{\sigma }
_{6})\,m_{0}^{2}+A_{t}\,{\sigma }_{5}\,m_{0}\,m_{12}+m_{12}^{2}\,({\sigma } 
_{4}-\sigma _{1})\,+\delta \,g^{2}\,v^{2}/16  
\nonumber \\ 
\gamma _{3} &\equiv &2\,(1-{\sigma }_{7}+A_{t}^{2}\,{\sigma }
_{6})\,m_{0}^{2}-A_{t}\,{\sigma }_{5}\,m_{12}\,m_{0} 
\nonumber\\
\gamma_4&=&
2 \,\gamma_2 \,\cos 2\beta+4\gamma_1\,\sin 2\beta
-(4+\delta)\,(g^2 \,v^2/8)\,\cos4\beta
\end{eqnarray}
%
The contributions $\Delta _{\zeta _{i}^{2}}$ are proportional to $%
\zeta _{i}$ so, for small enough changes from the MSSM case, the fine-tuning 
introduced with respect to these new parameters is small and sub-leading 
relative to that for the other parameters.

It is convenient to treat $\beta $ as the free 
parameter rather than $B_{0}$. 
Using the second minimum condition of (\ref{mincond}) 
(after replacing $m_{3}^{2}$ by (\ref{sm})), one finds
\begin{eqnarray} 
B_{0} &=&\frac{-1}{m_{0}\,\mu _{0}\,{\sigma }_{8}}\bigg\{\mu 
_{0}\,m_{12}\sigma _{2}\,+\mu _{0}\,m_{0}\,A_{t}\,\sigma _{3}-\frac{1}{2}(%
\tilde{m}_{1}^{2}+\tilde{m}_{2}^{2})\Big[\sin 2\beta   \nonumber  \label{B0} 
\\
&+&\frac{v^{2}}{\tilde{m}_{1}^{2}+\tilde{m}_{2}^{2}}\Big(\zeta _{1}\,(1+\sin 
^{2}2\beta )+\frac{\zeta _{2}}{2}\,\sin 2\beta +\delta \,(g^{2}/8)\,\sin 
2\beta \sin ^{2}\beta \Big)\Big]\bigg\}\quad  
\end{eqnarray} 

\medskip\noindent 
Note that  $\gamma _{1,4}$ brings some extra $\zeta _{1,2}$ 
dependence through $B_{0}$, while $\gamma _{2,3}$ are $\zeta _{1,2}$ 
independent. $\Delta _{p}$, $p=\{\mu 
_{0}^{2},m_{0}^{2},A_{t}^{2},B_{0}^{2},m_{12}^{2}\}$
 contain some $\mathcal{O}(\zeta _{i}^{2})$ terms,
 although the potential is only linear in $\zeta 
_{i}$. In the above expressions all coefficients $\sigma _{1,2,...,8}$ are 
evaluated at $m_{Z}$ and their values are given in (\ref{coefficients}). 
They depend only on the top Yukawa coupling at $m_{Z}$. The only 
approximation in obtaining the above expressions for $\Delta _{p}$'s is that 
we did not include the effect of derivatives (with respect to parameter $p$) 
acting on $\delta $ (the 
radiative correction to the quartic term).  
This is a legitimate approximation since  this effect is
numerically very small (for the MSSM alone it induces an error for 
fine-tuning $\Delta$ equal to or less than unity, while in 
the MSSM$_5$ the  error is even smaller ($1 \% $); 
for larger  $\tan\beta$ this error is further reduced).

The above results for the  fine-tuning measure simplify in the limit
 of ignoring the RG effects on the masses 
i.e. $\sigma _{1,2,3,...,6}=0;\,{\sigma }_{7,8}=1$.
In this case
\bea
\gamma _{1}&=& \mu _{0}\,B_{0}\,m_{0},\qquad 
\gamma _{2}=\delta \,g^{2}v^{2}/16,\,\,\qquad 
\gamma _{3}= 0,  
\nonumber\\
\gamma_4&=& 4\,B_0\,m_0\,\mu_0\,\sin 2\beta
+(g^2\,v^2/8)\,[\,\delta \cos 2\beta-(4+\delta)\,\cos 4\beta\,]
\eea
Ignoring RG effects on the quartic couplings too, $\delta=0$, 
then  $\gamma_4=4 B_0\,m_0\,\mu_0\sin 2\beta
-(g^2\,v^2/2)\,\cos4\beta$, $\gamma_2=0$.
Finally, 
in the limit $\zeta _{1,2}=0$ of the fine tuning relations $\Delta_p$,
one obtains analytical expressions for the EW 
scale fine tuning in the MSSM alone, with $\gamma_{1,...4}$ 
as in 
(\ref{gammaall}). Since these may be useful for other 
studies, they are provided in Appendix~\ref{appendixA}.

\subsection{The large $\tan\protect\beta$ limit.} 
 \label{largetanbeta} 
 
The above formulae for the fine-tuning simplify considerably in the limit of 
large $\tan \beta $. Ignoring terms suppressed by inverse powers of 
$\tan\beta $ one has
\medskip
\begin{eqnarray}
\Delta _{\mu_{0}^{2}} &=&\frac{-2\mu _{0}^{2}\,{\sigma }_{8}^{2}}{(1+\delta 
)\,m_{Z}^{2}}  \nonumber 
\\[7pt] 
\Delta _{m_{0}^{2}} &=&\frac{-m_{0}}{(1+\delta
  )\,m_{Z}^{2}}\,\,
\bigg[2\,{\sigma }_{7}\,m_{0}+A_{t}\,({\sigma }_{5}\,m_{12}-2A_{t}\,{\sigma }
_{6}\,m_{0})+\frac{\mu _{0}}{m_{0}}\frac{\zeta _{1}\,v^{2}\,m_{12}\,\sigma 
_{2}}{\gamma_{2}+(1+\delta /4)\,m_{Z}^{2}}\bigg]  
\nonumber\\[7pt]
\Delta _{m_{12}}^{2} &=&\frac{-m_{12}}{(1+\delta )\,m_{Z}^{2}}\,\,\bigg[
A_{t}\,{\sigma }_{5}\,m_{0}+2\,{\sigma }_{4}\,m_{12}-\frac{\zeta 
_{1}\,v^{2}\,\mu _{0}\,\sigma _{2}}{\gamma _{2}+(1+\delta /4)\,m_{Z}^{2}}
\bigg]  
\nonumber\\[0.07in] 
\Delta _{A_{t}^{2}} &=&\frac{A_{t}\,m_{0}}{(1+\delta )\,m_{Z}^{2}}\,\,
\bigg[
2\,A_{t}\,{\sigma }_{6}\,m_{0}-{\sigma }_{5}\,m_{12}+\frac{\zeta 
_{1}\,v^{2}\,\mu _{0}\,\sigma _{3}}{\gamma _{2}+(1+\delta /4)\,m_{Z}^{2}}\,
\bigg]\qquad\qquad\qquad\quad  
\nonumber\\[7pt] 
\Delta _{B_{0}^{2}} &=&\frac{-\zeta _{1}\,(m_{12}\,\sigma 
_{2}+A_{t}\,m_{0}\,\sigma _{3})\,\mu _{0}\,v^{2}}{(1+\delta 
)\,m_{Z}^{2}\,(\gamma _{2}+m_{Z}^{2}\,(1+\delta /4))}  
\nonumber\\[10pt]
\Delta _{\zeta _{i}^{2}} &=&0,\text{ \ }i=1,2  
\label{five} 
\end{eqnarray}

\medskip\noindent
One may see that all the fine tuning measures
are suppressed by the factor $(1+\delta )^{-1}$ 
demonstrating why quantum corrections to the quartic 
Higgs coupling can significantly reduce the fine-tuning. 
A similar effect applies at small $\tan \beta $ as well.

\subsubsection{Dimension-six operators} 
 
The operator analysis used here has a limited range of validity 
because it corresponds to integrating out new heavy degrees of freedom. If 
the mass of these degrees of freedom is not much above the energies being 
probed, the operator analysis breaks down and one must deal with the new 
degrees of freedom directly. The mass, $M_{\ast },$ at which this 
happens corresponds to the point where high dimension operators are not 
suppressed relative to low dimension operators. A measure of this may be 
obtained by dimensional analysis in which the operator matrix elements are 
taken to be determined by the energy scale being probed. Applied here, this 
implies that the operator analysis is reliable provided 
$\frac{m_{0},\mu }{M_{\ast }}\ll 1.$

A potential fault in this  estimate of  the range of 
convergence occurs because higher dimension operators 
may have anomalously large 
matrix elements. An example of this occurs for the dimension-six operators 
listed in Section~\ref{s1}. Consider the first dimension-six operator 
in eq.(\ref{dim6}) 
\medskip
\begin{equation} 
\mathcal{O}_{2}\supset \frac{c_{1}}{M_{\ast }^{2}}\Big[\,S^{\dagger 
}S(H_{2}^{\dagger }\,\exp V\,H_{2})^{2}\Big]_{D}\supset 
\frac{c_{1}\,m_{0}^{2}}{M_{\ast }^{2}}\,|h_{2}|^{4},
\,\,\,\,\,(c_{1}\sim \mathcal{O}(1)).  
\label{six} 
\end{equation}

\bigskip\noindent
This should be compared to the leading quartic Higgs term coming 
from the dimension-five operators in eq.(\ref{scalarV})
 that contributes at 
 $\cO\left( \frac{2\mu _{0}}{M_{\ast }}\left\vert h_{2}\right\vert 
 ^{2}h_{1}h_{2}\right)$.
One may see that the relative magnitude of 
the dimension-six to dimension-five contributions is 
$\cO\left( \frac{m_{0}^{2}}{2 \mu _{0}M_{\ast }}\tan \beta \right) .$
 Thus, strictly, 
the region of validity of the dimension-five operator analysis is 
$\frac{m_{0}^{2}}{2\mu _{0}M_{\ast }}\tan \beta \ll 1.$
However, as discussed in the next section, the new physics
 generating this dimension-six 
operator is different from that generating the dimension-five operator and 
so their coefficients should be uncorrelated. In this case the addition of 
higher dimension operators can reduce the fine tuning for some region in 
parameter space
 so that the analysis with dimension-five 
operators only will provide a useful upper bound even in regions where 
dimension-six contributions are significant.
For this reason the region of validity of the dimension-five operators 
analysis is better described by the original $m_0/M_*\ll 1$ and
$\mu_0/M_*\ll 1$ condition.
This keeps the corrections coming from operators with correlated 
coefficients small. To see this more explicitly consider 
the effect of the term in eq.(\ref{six}) on fine-tuning. Writing 
$\nu \equiv c_{1}\,m_{0}^{2}/M_{\ast }^{2}$
one has 
\bigskip 
\bea
\Delta _{p}\approx -\frac{2\,p}{(1+\delta +8\nu /g^{2})\,m_{Z}^{2}}
\frac{\partial \tilde{m}_{2}^{2}}{\partial p} 
\eea

\bigskip\noindent
where, for example, $p$ can be $\mu _{0}^{2}$.
The partial derivative is readily obtained from eq.(\ref{sm}).  The dominant 
effect of the d=6 operator on fine tuning is the appearance 
of the effect of the  $8\nu /g^{2}$
term in the denominator, reducing the fine tuning for the appropriate
sign of $8\nu /g^{2}$.
(note that c.f. eq.(\ref{five})  such a 
term is not generated by the dimension-five term at large $\tan\beta$).
The effect of this reduction is sizeable.
 Taking, for example, 
$m_{0}/M_{\ast }\approx 1/10$  and $c_{1}=3$, then 
$8\nu /g^{2}\approx 1/2$ which is close to the numerical value of 
$\delta $ entering the denominator. One sees that $d=6$
operators can bring a reduction of $\Delta _{\mu _{0}^{2}}$
relative to that of the MSSM including top/stop effects, of order 
$(8\nu/g^2)/(1+\delta +8\nu /g^{2})\approx 30\%$.

In the following numerical analysis we  include only the effects of
dimension-five operators. The convergence criterion found above
gave $m_0/M_*\ll 1$ and $\mu_0/M_*\ll 1$. 
In our following  numerical analysis this bound is comfortably satisfied 
when we take $m_0/M_*, \,\mu_0/M_*\leq 0.035$,
giving upper values $\zeta_{1,2}\leq 0.07$.

\subsection{Numerical results}\label{numerics} 
 
We are now in a position to determine the fine-tuning in the 
extended MSSM Higgs sector. We will present this as a function of
the mass $m_{h}$  of the lightest CP even Higgs. This is 
given by:
\begin{eqnarray} \label{mh2}
m_{h}^{2} &=&\frac{1}{2}\Big[m_{A}^{2}+m_{Z}^{2}-\sqrt{w}+\xi \Big]  
\nonumber  \label{finalmh} \\[8pt] 
&+&{\zeta _{1} {\ v}^{2}\,\sin 2\beta }\,
\bigg[1+\frac{m_{A}^{2}+m_{Z}^{2}}{
\sqrt{w}}\bigg]+\frac{\zeta _{2} {\ v}^{2}}{2}\,\bigg[1-
\frac{(m_{A}^{2}-m_{Z}^{2})\,\cos ^{2}2\beta }{\sqrt{w}}\bigg] 
\end{eqnarray} 
where 
\begin{eqnarray} 
w &\equiv &[(m_{A}^{2}-m_{Z}^{2})\,\cos 2\beta +\xi ]^{2}+\sin ^{2}2\beta 
\,(m_{A}^{2}+m_{Z}^{2})^{2}  \nonumber  \label{mhf} \\[9pt] 
m_{A}^{2} &=&\tilde{m}_{1}^{2}+\tilde{m}_{2}^{2}+\xi /2+\zeta 
_{1}\,v^{2}\,\sin 2\beta -(1/2)\,\zeta _{2}\,v^{2};\quad 
\xi \equiv \delta m_{Z}^{2}\sin ^{2}\beta
\end{eqnarray}

\begin{figure}[t] 
\begin{tabular}{cc|cr|} 
\parbox{7cm}{\psfig{figure=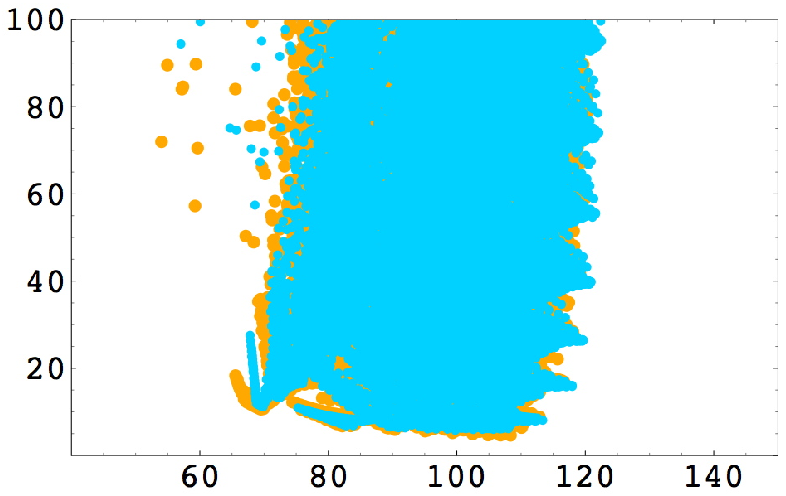,  
height=5.cm,width=6.7cm}} \hspace{0.5cm}  
\parbox{7cm}{\psfig{figure=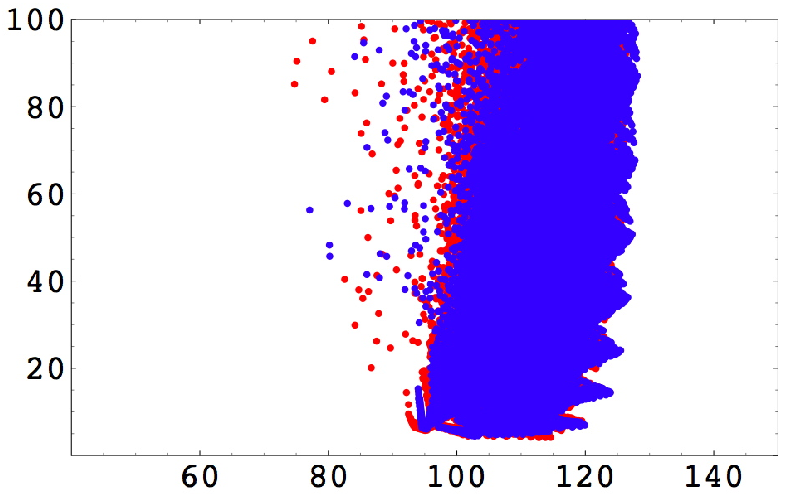, 
height=5.cm,width=6.7cm}}
\end{tabular}%
\def\baselinestretch{1.1}
\caption{{\protect\small Left figure (a): the MSSM fine tuning
    $\Delta$ as a function of $m_{h}$;
 Right figure (b): the fine tuning in the MSSM with $d=5$ 
operators in terms of $m_{h}$, with 
$\zeta _{1}\!=\zeta_{2}\!=\!0.03$.
 In both figures, the top pole mass considered is $m_{t}=174$ GeV 
for blue (dark blue) 
areas and $m_{t}=171.2$ for yellow (red) areas, respectively. 
Larger $m_{t}$ 
input (blue) shifts the plots towards higher $m_{h}$ by 2-5 GeV. In both 
figures the parameters space scanned 
is: $1.5\leq \tan \protect\beta \leq 10$,
 $50\,\mathrm{GeV}\!\leq m_{0},m_{12}\leq\! 1$ TeV,
$130\,\mathrm{GeV}\leq  
\protect\mu _{0}\leq 1$ TeV, $-10\leq A_{t}\leq 10$. }} 
\label{fig1ab} 
\end{figure} 
\begin{figure}[th] 
\begin{tabular}{cc|cr|} 
\parbox{7.cm}{\psfig{figure=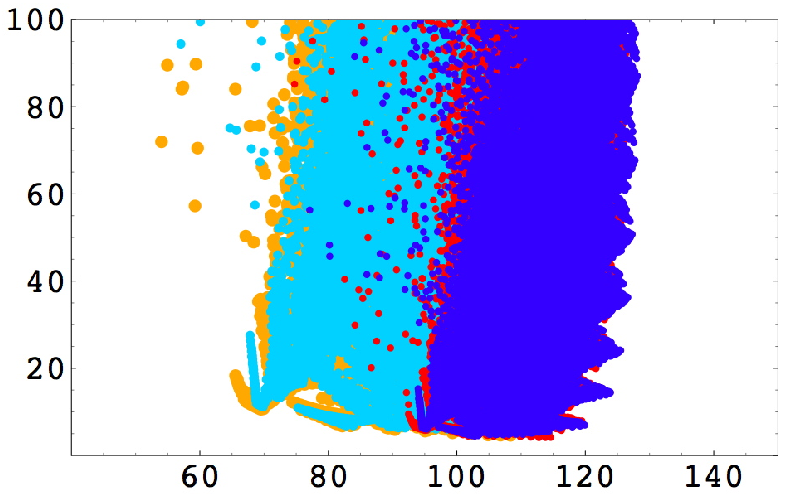, 
height=5.cm,width=6.7cm}} \hspace{0.5cm}  
\parbox{7.cm}{\psfig{figure=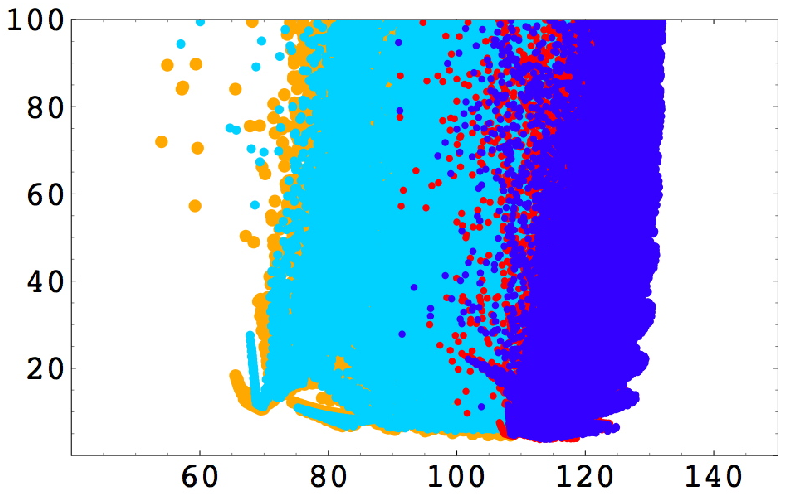, 
    height=5.cm,width=6.7cm}} 
\end{tabular}
\def\baselinestretch{1.1}
\caption{{\protect\small Left figure (a): the fine tuning $\Delta $ as a
function of $m_{h}$. 
$\Delta $ of MSSM is plotted in light blue ($m_{t}=174$ GeV) with an orange 
edge ($m_{t}=171.2$); $\Delta $ of MSSM with d=5 operators with $\protect%
\zeta _{1}=\protect\zeta _{2}=0.03$ is plotted in dark blue ($m_{t}=174$ 
GeV) with a red edge ($m_{t}=171.2$). Right figure (b): similar to figure 
(a) but with $\protect\zeta _{1}=\protect\zeta _{2}=0.05$. Non-zero or 
larger $\protect\zeta _{i}$ (dark blue and red areas) shift the plots to 
higher $m_{h}$ to allow a reduced $\Delta $ for higher $m_{h}$. In both 
figures $1.5\leq \tan \protect\beta \leq 10$; $50\,\mathrm{GeV}\leq 
m_{0},m_{12}\leq 1$ TeV, $130\,\mathrm{GeV}\leq \protect\mu _{0}\leq 1$ TeV,  
$-10\leq A_{t}\leq 10$.}} 
\label{fig2ab} 
\end{figure}

\medskip\noindent
Using the results of the previous section we compute the fine tuning 
for a sample of points in parameter space in the region with:
$1.5\leq \tan \beta 
\leq 10$, $50\,\mathrm{GeV}\leq m_{0},m_{12}\leq 1$ TeV,
 $130$~GeV~$\leq \mu_0\leq 1$ TeV,
$-10\leq A_{t}\leq 10$  and $171.2\leq m_{t}\leq 174$ GeV 
consistent with $m_{t}=172.6\pm 1.4$ GeV \cite{TeV}, 
and with the signs
for $\zeta_{1,2}$ chosen so as to reduce the fine tuning.

\begin{figure}[t] 
\begin{tabular}{cc|cr|} 
\parbox{7.cm}{ 
\psfig{figure=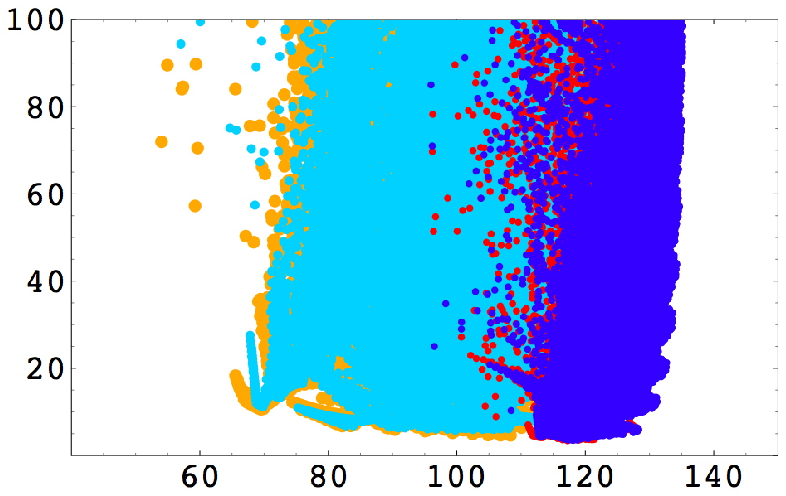,  
height=5.cm,width=6.7cm}} \hspace{0.5cm}  
\parbox{7.cm}{ 
\psfig{figure=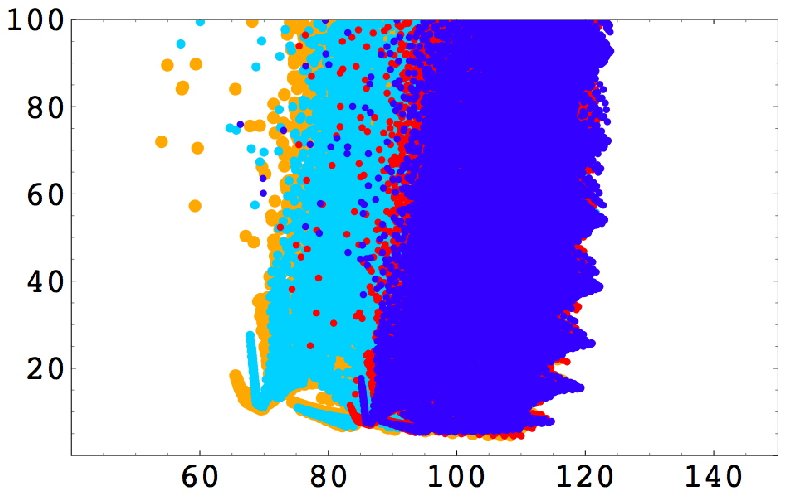, 
height=5.cm,width=6.7cm}} 
\end{tabular}
\def\baselinestretch{1.1}
\caption{{\protect\small As in Fig.\protect\ref{fig2ab} but with: left 
figure (a): $\zeta_1=0.07$, $\zeta_2=0$; right figure (b):
 $\zeta_1=0$, $\zeta_2=0.1$}} 
\label{fig4ab} 
\end{figure}

The results
 are shown in Figures~\ref{fig1ab} to \ref{fig4ab}. Note that in these
 figures the structure apparent at small $\Delta$ and large $m_h$  is
 probably a scanning artefact. We expect the under-dense wedge shaped
 regions will be filled in with a more dense parameter sample.
 Similarly at very large $\Delta$, corresponding to very precise
 relationships between parameters, there will be some points
 corresponding to high values of $m_h$ that are not picked
 up by our finite parameter scan.

 Turning to our results, as a benchmark
 Figure 1(a) shows the EW scale fine 
tuning $\Delta $ of eq.(\ref{ft})
 of the MSSM as a function of $m_{h}$.
 One may see 
that $\Delta \geq 18$ for values $m_{h}\geq 114.4$ GeV, 
the current LEPII bound. Figure 1(b) shows $\Delta $ for the case 
of the MSSM with dimension-five operators added, with $\zeta _{1}=\zeta 
_{2}=0.03$. The dominant effects in Figure~\ref{fig1ab}(b) are mostly due to 
the effect of non-zero $\zeta _{1}$, which comes from the 
supersymmetric part of the higher dimensional operator. One may see a 
systematic shift of the allowed region to higher $m_{h}$  which 
(for positive $\zeta _{i})$  is driven by an increase in the 
quartic Higgs coupling which appears in the denominator of the fine tuning 
measure ($c.f.$ eq.(\ref{five})). 
The overall result is that the 
minimum amount of fine-tuning $\Delta $ in the 
presence of $d=5$ effective operators 
is small, of order $\Delta \approx 6$, for $m_{h}$ from 
$95$ to $119$ GeV. Therefore non-zero $\zeta _{i}$ can accommodate larger $%
m_{h}$ while keeping a $\Delta $ significantly smaller than in the MSSM. To 
illustrate the change more directly we superpose both plots in 
Figure~\ref{fig2ab} (a). The effect is enhanced for larger operator
coefficients, as 
may be seen in Figure~\ref{fig2ab} (b), where $\Delta $ is presented for
$\zeta_{1}=\zeta _{2}=0.05$, again shown relative to the MSSM case. One can see 
that in this case values of $m_{h}>114.4$ GeV can have a low $\Delta \approx 
6$. Therefore $\Delta$ can be significantly reduced from 
the MSSM case, for a similar $m_{h}$. This conclusion is further supported 
by the plots in Figure~\ref{fig4ab} where other values for $%
\zeta _{i}$ are considered. From all plots shown one sees that $\Delta <10$ 
is easily satisfied for values of the Higgs mass that can be as large
as $130$ GeV, depending on the exact values of $\zeta_{i}$. 

Note that, in the MSSM, 
$\Delta$ increases for low $\tan\beta$\,\, ($\ll 10$)
 and $m_h$ above the LEPII bound.
However, c.f. eq.(\ref{mh2}),
the effect of the $d=5$ operators is important 
for low $\tan\beta$ and  in their presence
$\Delta$ actually decreases for low $\tan\beta$. 
Thus the reduction in the fine tuning at very low
$\tan\beta$ relative to the MSSM case
is much more  marked than that shown.

The lower amount of fine tuning in the presence of effective 
operators is due to two effects. The first, already mentioned, is the 
presence of additional quartic Higgs couplings enhancing the denominator 
which determines the Higgs via $v^{2}=-m^{2}/\lambda $  thus 
allowing for a smaller electroweak breaking scale. The second is the fact 
that higher dimensional operators add a tree level contribution to the Higgs 
mass, which reduces the need for large quantum contributions, and therefore 
the fine tuning. 
 
What is the scale of new physics needed for this reduction in fine 
tuning? Using eq.(\ref{coeffs}) we find that the scale of new physics 
is
\begin{equation}
M_*\approx 2\mu_0/\zeta_1 \approx 
(40\,\,\,\mathrm{to}\,\,\,65) \times \mu_0,\qquad 
\zeta_{1,2}=0.05\,\,\,\mathrm{to}\,\,\,0.03
\end{equation}
With $\mu_0$  between the EW scale and 1 TeV, 
this shows that large values of $M_*$ are allowed: $M_*\approx (5.2\,\,%
\mathrm{to} \,\,8.45)$ TeV for $\mu_0=130$ GeV and $M_*\approx (8\,\,\mathrm{%
to}\,\,13)$ TeV for $\mu_0=200$ GeV. For larger $\mu_0$ one obtains
values of $M_*$ above the LHC reach. Finally, for $\zeta_1=0.07$ but with
$\zeta_2=0$, 
one has $M_*\approx 30\times \mu_0$ and  $\Delta<10$ for $m_h\approx 130$ GeV.
Thus, the
 EW fine tuning is small $\Delta<10$ for $114\leq m_h\leq 130$ GeV,
for rather conservative values of $\zeta_{1,2}$.
To relax these values one can use that an increase of $\zeta_1$
by $0.01$ increases  $m_h$ by $2$ to $4$ GeV for the same $\Delta$.

\subsection{The origin of ``new physics''}\label{origin}
 
The presence of a higher dimension operator signals new physics and 
it is important to ask what this new physics can be. In the context of new 
renormalisable interactions it may come from the effects of new chiral 
superfields or from new gauge vector superfields. Consider chiral 
superfields first. One may readily obtain the $d=5$  operator of 
eq.(\ref{dimensionfive}) by integrating out a gauge singlet or a
triplet 
\cite{Dine}. Consider the case of a massive gauge singlet 
$X$ with Lagrangian 
\medskip
\[ 
\mathcal{L}_{X}=\int d^{4}\theta \,X^{\dagger }X+\bigg\{\int d^{2}\theta \,%
\Big[\mu H_{1}H_{2}+\lambda _{x}\,X\,H_{1}H_{2}+\frac{1}{2}\,M_{\ast }\,X^{2}%
\Big]+h.c.\bigg\}. 
\] 

\bigskip\noindent
For $M_{\ast }\gg \mu ,$ $m_{0}$, one may use the eqs of 
motion to integrate out $X$, giving, to leading order in inverse 
powers of $M_{\ast },$ 
\begin{equation} 
\mathcal{L}_{X}^{effective}=\frac{-\lambda _{x}^2}{2\,M_{\ast }}\int 
d^{2}\theta \,(H_{1}H_{2})^{2}+h.c.  \label{ops} 
\end{equation} 
 
\medskip\noindent
The supersymmetry breaking terms associated with this 
operator are obtained by replacing $\lambda \rightarrow \lambda (S)$
giving the $d=5$ operator of interest.
Note that $\mathcal{L}_{X}$ has a similar form to that of 
the NMSSM. However in the NMSSM the singlet field has mass of order the 
electroweak breaking scale and cannot be integrated out whereas here we are 
taking the singlet mass to be much larger than the EW scale. 
 
However, the origin of the $d=5$ operator cannot be 
uniquely ascribed to a gauge singlet field. Indeed it may equally well point 
to the existence of $SU(2)$ triplets \cite{Espinosa:1998re,
Espinosa:1998re,Espinosa:1991gr,Espinosa:1991wt}
$T_{1,2,3}$  of hypercharge $\pm 1,0$. In this case a 
Lagrangian of the form 
\medskip  
\[ 
\mathcal{L}_{T}=\!\!\int \!d^{4}\theta \,\Big[T_{1}^{\dagger 
}e^{V}T_{1}+T_{2}^{\dagger }e^{V}T_{2}\Big]+\!\int d^{2}\theta \,\Big[\mu 
H_{1}H_{2}\!+\!M_{\ast }T_{1}T_{2}\!+\!\lambda 
_{1}H_{1}T_{1}H_{1}\!+\!\lambda _{2}H_{2}T_{2}H_{2}\Big]\!+\!h.c  
\]

\medskip\noindent
gives, to lowest order in $1/M_{\ast }$, 
eq.(\ref{ops}) with $\lambda _{x}^2$ replaced by $\lambda _{1}\lambda _{2}$.
 More generally, one can generate the $d=5$
operator through a combination of both gauge singlets and triplets. 
However  note that the pure singlet $X$  case has the advantage of not 
affecting the gauge couplings unification (at one-loop), 
which is not true for the $SU(2)$ triplet. 
 
What about additional, massive, $SU(2)$ doublets that 
couple to the MSSM Higgs sector? One may readily show that integrating them 
out does not generate, to lowest order in $1/M_{\ast }$, an 
operator of the type (\ref{ops}).
 
There remains the possibility that the new physics is due to the 
effect of new massive vector gauge superfields. The simplest example is the 
case there is a new $U(1)^{\prime }$ gauge symmetry under which 
the Higgs sector is charged. This brings extra quartic contributions to the 
scalar potential that are expected to reduce the fine-tuning 
\cite{Batra:2003nj,kaplan,Bellazzini:2009ix}. Assuming the
$U(1)^{\prime }$ 
 is broken at $M_{\ast }$  one obtains the effective 
Lagrangian to leading order in inverse powers of $M_{\ast }$
 given by
\medskip
\[ 
\mathcal{L}_{U(1)}^{effective}=-\frac{g^{^{\prime }2}}{M_{\ast }^{2}}\int 
d^{4}\theta \,\,\Big[q_{1}H_{1}^{\dagger }e^{V}H_{1}+q_{2}H_{2}^{\dagger 
}\,e^{V}H_{2}\Big]^{2} 
\]

\medskip\noindent
where $g^{\prime }$ is the $U(1)^{\prime }$
coupling and $q_{1,2}$ are the charges of the Higgses under 
$U(1)^{\prime }$ ($q_{1}+q_{2}=0$). Note that, after 
including the associated supersymmetry breaking operators, this corresponds 
to the $d=6$ effective operators \cite{Dine} of eq.(\ref{dim6}) and 
that no $d=5$ operators are generated. 

\medskip  
In summary, the requirement that the SUSY extension of the MSSM 
should not have significant fine tuning may indicate the presence of
the 
$d=5$  operator of eq.(\ref{dimensionfive}) which, in turn, suggests 
the presence of a massive gauge singlet and/or a $SU(2)$  triplet. 
This is the simplest interpretation based on new renormalisable interactions 
but other, more complicated possibilities to generate the 
$d=5$ operator may be 
possible.

\bigskip
\subsection{Further remarks on fine tuning}

Effective field theory approaches to the fine tuning 
of the electroweak scale were used before in models of low susy breaking
scale scenarios \cite{Casas:2003jx} where both $d=5$ and $d=6$
operators were included. The model in \cite{Casas:2003jx} introduces
supersymmetry breaking through coupling of MSSM states to a 
SM singlet field responsible for supersymmetry breaking.
After integrating this field out, in addition to the $d=5$ operator 
considered here, there are correlated contributions from the 
$d=6$ operators. Using this, the authors find the fine tuning can be
very small even for an arbitrarily high Higgs mass, provided the
scale of supersymmetry breaking is less than $500$ GeV.

How does this analysis relate to the one presented here?
 The examples given in  \cite{Casas:2003jx} are found varying the ratio $\tilde
 m/M$ in the range $0.05$ to $0.8$ where $\tilde m$ is the
 supersymmetry breaking scale and $M$ is the messenger mass.
For $\tilde m/M$ small, the fine tuning is close to that in the MSSM
 but reduces rapidly for $\tilde m/M$ large; in this latter case the
 fine tuning actually reduces as the Higgs mass increases. This range
of values for $\tilde m/M$ corresponds to a 
choice of our $m_0/M_*$ and $\mu_0/M_*$ in a similar range.
The upper value strongly violates our criterion for applicability of
the operator analysis and is a factor of $\approx 10$ larger than the value
 chosen in Figure~\ref{fig4ab}(a).
Ignoring, for the moment, the fact that the contributions of
 higher dimension operators 
are expected to be large for this choice of mediator mass, we can ask 
what this choice of mediator mass in our analysis would give for 
the Higgs mass consistent with small 
$\Delta$. Since the change in our upper bound on the Higgs mass
 roughly scales with the coefficient of the $d=5$ operator
 (eq.(\ref{mh2})), this would allow a Higgs mass in the region of
 $276$ GeV, much larger than our earlier conservative estimates.
 However, as we have stressed, for this value of the messenger mass the operator
analysis breaks down  and one should do the analysis including the
 messenger fields explicitly.

 \bigskip
\section{Conclusions} 
 \label{conclusions} 
 
The LEPII lower bound on the Higgs mass places MSSM Higgs physics at
the forefront  of supersymmetry phenomenology. While this bound can be 
satisfied by including the MSSM quantum corrections, it 
(re)introduces some amount of fine tuning in the model.
To reduce the fine tuning may require new physics beyond the MSSM 
which can be parametrised by higher dimensional operators.
In this paper we used an effective field 
theory framework with $d=5, 6$ operators in the MSSM Higgs sector, 
and  presented a  model independent approach to the fine tuning problem.

We obtained exact  analytical results for the EW scale fine tuning
in the MSSM with dimension-five operators,  which are also applicable
to the pure MSSM case in the limit the coefficients of the 
higher dimension operators vanish. This calculation  included 
one-loop corrections  to the soft masses and dominant top
Yukawa effects on the quartic terms of  the potential.
Similar analytical results  were given for a general two-Higgs 
doublet model.

Fine tuning proves to be  very sensitive to the addition of higher
dimensional operators and this is  mostly due to extra
corrections to the quartic couplings of the Higgs field.
For the case of dimension-five operators 
we showed that one can maintain a reduced fine-tuning
 $\Delta<10$ for a Higgs mass above the LEPII bound and
 as large as $m_h\approx 130$ GeV, 
for  the parameter space considered, with low $\tan\beta$ ($\tan\beta<10$). 
 The scale of new physics $M_*$  responsible for the reduction in fine 
tuning can be rather large, for example $M_*\approx 2\mu_0/\zeta_1 \approx 
(40\,\mathrm{to}\,65) \times \mu_0$, for $\zeta_{1,2}=0.05\,\,\mathrm{to}
\,\,0.03 $, and $M_*\approx 30 \times \mu_0$ for $\zeta_1=0.07$, $\zeta_2=0$.
For values of $\mu_0$  between the electroweak  scale and 1 TeV, 
these results show that large values of $M_*$ are allowed; in the former case
$M_*\approx~(5.2\,\,\mathrm{to} \,\,8.45)$ TeV for $\mu_0=130$
 GeV and $M_*\approx (8\,\,\mathrm{to}\,\,13)$
 TeV for $\mu_0=200$ GeV. For larger $\mu_0$, larger
 values of $M_*$ are possible, even above the LHC reach.
These results follow from  rather conservative choices for
 the coefficients of the quartic couplings  induced by 
the dimension-five  operators, to ensure the convergence of the 
effective operator expansion.

Our numerical analysis included the effect of dimension-five operators
only. These  give the leading corrections at low $\tan\beta$, being 
proportional to $1/M_*$. However, dimension-six operators, suppressed
by $m_0^2/M_*^2$ or $\mu_0^2/M_*^2$, give contributions that can be
 enhanced by large $\tan\beta$;
 for $(\tan\beta\,m_0)/M_*>1$ or $(\tan\beta\,\mu_0)/M_*>1$
these will be the leading terms. For this reason they should be
included at large $\tan\beta$ and we hope to extend our analysis to
the dimension-six case in the future.

Of course the crucial question is what is the origin of the physics
beyond the MSSM giving rise to these operators?
The dimension-five operator can be generated by a gauge singlet
superfield or a $SU(2)$ triplet superfield of mass of $\cO(M_*)$
 coupling to the Higgs sector. The dimension-six operators can be
generated, for example, by an extra gauge symmetry with a massive gauge
supermultiplet or additional (Higgs-like) $SU(2)$ doublet supermultiplets
of mass $\cO(M_*)$. If the fine tuning criterion is indeed of physical
relevance, the significant amount of fine tuning found in the MSSM
already indicates the need for such additional degrees of freedom.

\vspace{1.5cm}
\section*{\textbf{Acknowledgements: }} 
 
This work was partly supported by the European Union
Research and Training network  contract MRTN-CT-2006-035863.
SC is supported by the UK Science and Technology Facilities Council 
(PPA/S/S/2006/04503).
 D.G. thanks I. Antoniadis, E. Dudas and P. Tziveloglou for 
many interesting discussions related to this topic.

\newpage
\section*{Appendix}
\def\theequation{A-\arabic{equation}}
\def\thesubsection{A}
\setcounter{equation}{0}
\subsection{Fine tuning expressions in the MSSM.}
\label{appendixA}

The coefficients $\sigma_i$ used in the text, Section~\ref{general} are:
\medskip
\bea\label{coefficients}
\sigma_1(t_z)&=&0.532,  \qquad \qquad\qquad \qquad\qquad \qquad
\sigma_2(t_z)=0.282\,(4.127\,h_t^2-2.783)(1.310-h_t^2)^{1/4}\nonumber\\
\sigma_3(t_z)&=& -0.501\,h_t^2\,(1.310-h_t^2)^{1/4}, \qquad \quad\,\,
\si_4(t_z)=0.532-5.233\,h_t^2+1.569\,h_t^4\nonumber\\
\si_5(t_z)&=& 0.125\,h_t^2\,(10.852\,h_t^2-14.221), \qquad 
\si_6(t_z)= -0.027\,h_t^2\,(10.852\,h_t^2-14.221)\nonumber\\
\si_7(t_z)&=&1-1.145 \,h_t^2,  \qquad \qquad\qquad \qquad\quad\,
\si_8(t_z)= 1.314 \,(1.310-h_t^2)^{1/4}
\eea
\medskip\noindent
where $h_t$ is evaluated at $m_Z$ and
$m_t=h_t(t_{m_t})\,(v/\sqrt 2)\sin\beta$.  

\vspace{1cm}
 In the MSSM one obtains the following  analytical 
expressions for  fine-tuning
(these are obtained from the results in Section~\ref{general} by setting
 $\zeta_{1,2}=0$):
\bigskip
\bea
\Delta_{\mu_0^2}
&=& -\frac{1}{v^2\,D}\,
\Big\{(g^2 v^2/8)\,\gamma_1\,\sin 4\beta \big[\,\delta
-  (4+\delta)\,\cos 2\beta\,\,\big]
+ 2\, \big[\,2\mu_0^2\,\si_8^2-\gamma_1 \sin2\beta\,\big]
\gamma_4
\Big\}\qquad\quad
\eea
\smallskip
\bea
\Delta_{m_0^2}&=& - \frac{1}{4\,v^2\,D}
\Big\{
 2\,v^2\Big[2\,(\gamma_1\!-\mu_0\,m_{12}\sigma_2)
\cos 2\beta\!+\gamma_3\sin
  2\beta\Big]
\Big[\delta g^2\cos\beta\sin^3\!\beta
-(g^2/2)\,\sin4\beta\Big]
\nonumber\\[7pt]
&+& 
8\,\gamma_4
\Big[2 m_0^2 -\gamma_3\,\sin^2\beta
+(m_{12}\,\mu_0\,\sigma_2-\gamma_1)\,\sin2\beta\Big]\Big\}
\eea
\smallskip
\bea
\!\Delta_{m_{12}^2}\!&=&\!\!\!
-\frac{m_{12}}{v^2\,D}\,\,
\Big\{\,\frac{g^2\,v^2}{2}\, 
\Big[ \, 2\mu_0\sigma_2\cos2\beta-
\big(A_t\,\si_5\,\,m_0+2\,m_{12}\,(\si_4-\sigma_1)\big)\,\sin 2\beta\Big]
\Big[ 
\delta \,\cos\beta\sin^3\beta
\nonumber\\[7pt]
&-&\!\!\! \frac{1}{2}\sin4\beta\Big]+\! 2\,
\Big[2 m_{12}\,\sigma_1-\mu_0 \sigma_2\sin 2\beta
+\big(A_t\,\si_5\,m_0+2\,m_{12}(\si_4-\sigma_1)\big)\sin^2\beta\Big]
\gamma_4
\Big\}
\eea
\smallskip
\bea
\Delta_{A_t^2}&=&
-\frac{A_t}{v^2\,D}\,\,\Big\{2\,m_0\sin
\beta\,\big[\,2\mu_0\,\sigma_3\,\cos\beta+(2\,A_t\,\si_6\,m_0
-\si_5\,m_{12})\sin\beta\,\big]\,(-\gamma_4)
\\[7pt]
&&\!\!\!\!\!\!\!\!\!\!\!\!\!\!
+\,\big[\,\delta\cos\beta\sin^3\beta
-\frac{1}{2}\sin4\beta\,\big]
\big[\,\mu_0\,\sigma_3\cos 2\beta- \si_5\,m_{12}/2\,\sin2\beta
+A_t\,\si_6\,m_0\sin 2\beta\big] m_0\, g^2 v^2
\Big\}\nonumber
\eea

\noindent
and
\medskip
\bea
\Delta_{B_0^2}=
-\frac{2 \,B_0\,m_0\,\mu_0\, \sigma_8}{v^2\,D}\,
\Big\{
(\delta g^2\,v^2/16-\gamma_2)\,\sin4\beta
-(4+\delta) \frac{g^2 v^2}{8}\sin^3 2\beta
-4\gamma_1\,\sin^2 2\beta \Big\}\,\,\,
\eea

\medskip\noindent
The denominator  $D$ is now
\medskip
\bea
D&=& 
\frac{1}{4}\,g^2\,\Big\{-
g^2\,v^2\,(\delta \cos\beta\sin^3\beta-1/2\,\sin4\beta)^2
- 2\,(\cos^2 2\beta+\delta\sin^4\beta)
(-\gamma_4)
\Big\}
\eea

\medskip\noindent
with the notation
\medskip
\bea\label{gamma}
\gamma_1&\equiv &\mu_0\,(B_0\,m_0\,\sigma_8
+m_{12}\,\sigma_2+A_t\,m_0\,\sigma_3)
\nonumber\\[7pt]
\gamma_2&\equiv & (-1+\si_7-A_t^2\,\si_6)\,m_0^2+A_t\,\si_5\,m_0\,m_{12}+
m_{12}^2\,(\si_4-\sigma_1)\,+\delta\,g^2\,v^2/16\nonumber\\[7pt]
\gamma_3&\equiv& 2\,(1-\si_7+A_t^2\,\si_6)\,m_0^2-A_t\,\si_5\,m_{12}\,m_0
\nonumber\\[7pt]
\gamma_4&\equiv &
4\gamma_1\sin 2\beta+2 \gamma_2 \cos
 2\beta- (4+\delta)\,(g^2\,v^2/8) \cos 4\beta
\eea

\medskip\noindent
and finally
\medskip
\bea
B_0=
\frac{1}{m_0\,\mu_0\,\sigma_8}
\bigg\{\,
\frac{1}{2}\,(\tilde m_1^2+\tilde m_2^2)\,
\sin2\beta \,\Big[
1+\frac{\delta\, g^2\,v^2/8}{\tilde m_1^2+\tilde m_2^2}
\sin^2\beta
\Big]
-\mu_0\,m_{12} \sigma_2\,-\mu_0\,m_0\,A_t\,\sigma_3
\bigg\}
\eea

\bigskip\noindent
The results for fine tuning given above considered a common
bare gaugino mass, but this restriction can easily be lifted
to obtain similar expressions.


\bigskip\bigskip\bigskip
\def\theequation{B-\arabic{equation}}
\def\thesubsection{B}
\setcounter{equation}{0}

\subsection{Evaluation of 
fine-tuning $\Delta_p$ in general two-Higgs doublet models.}\label{appendixB}

We present here the analytical result for the EW fine-tuning 
wrt a parameter $p$, for an arbitrary  two-Higgs doublet model. 
This can be immediately applied to 
a specific  model. Start with the general potential 
\medskip
\bea
V&=&\tilde m_1^2\,\vert H_1\,\vert^2
+\tilde m_2^2\,\vert H_2\,\vert^2
-(m_3^2\,H_1 \cdot H_2+h.c.)\nonumber\\[6pt]
&+&
\frac{1}{2}\,\lambda_1 \,\vert H_1\,\vert^4
+\frac{1}{2}\,\lambda_2 \,\vert H_2\,\vert^4
+\lambda_3 \,\vert H_1\,\vert^2\,\,\vert H_2\,\vert^2\,
+\lambda_4\,\vert\,H_1\cdot H_2\,\vert^2\nonumber\\[5pt]
&+&
\bigg[\,\,
\frac{1}{2}\,\lambda_5\,(H_1\cdot  H_2)^2+\lambda_6\,\vert\,H_1\,\vert^2\, 
(H_1 \cdot H_2)+
\lambda_7\,\vert\,H_2\,\vert^2\,(H_1 \cdot H_2)+h.c.\bigg]
\eea

\medskip\noindent
In the particular case of MSSM with $d=5$ operators
\medskip\bea
\lambda_1=\frac{1}{4}\,(g_2^2+g_1^{2}),\qquad
\lambda_2&=&\frac{1}{4}\,(g_2^2+g_1^{2})\,(1+\delta),\qquad
\lambda_3=\frac{1}{4}\,(g_2^2-g_1^{2})\nonumber\\[6pt]
\lambda_4= -\frac{1}{2}\,g_2^2,&&\qquad
\lambda_5=\zeta_2,\qquad
\lambda_6=\lambda_7=\zeta_1
\eea

\medskip\noindent
while in the MSSM alone one also sets $\zeta_1=\zeta_2=0$.

The minimum conditions can be written
\bigskip
\bea\label{min}
-v^2=\frac{m^2}{\lambda},\qquad 
2\,\lambda\,\frac{\partial
  m^2}{\partial\beta}-m^2\,\frac{\partial\lambda}{\partial \beta}
=0
\eea
with
\bea 
m^2&=&
\tilde m_1^2\,c_\beta^2+\tilde m_2^2\,s_\beta^2-m_3^2\,s_{2\beta} =
\tilde m_2^2- \frac{2}{u}\,m_3^2 +\frac{1}{u^2}
(\tilde m_1^2-\tilde m_2^2)+\cO(1/u^3)
\nonumber\\[6pt]
\lambda&=& \frac{\lambda_1}{2}\,c_\beta^4+\frac{\lambda_2}{2}
\,s_\beta^4
+(\lambda_3+\lambda_4+\lambda_5)\,s_\beta^2\,c_\beta^2
+2\lambda_6\,c_\beta^3\,s_\beta+ 2\lambda_7\,c_\beta\,s_\beta^3
\nonumber\\[6pt]
&=&
\frac{\lambda_2}{2}+\frac{2}{u}\,\lambda_7
+\frac{1}{u^2}\,(\lambda_3+\lambda_4+\lambda_5-\lambda_2)+\cO(1/u^3),
\eea

\bigskip
\noindent
with $s_\beta=\sin\beta$, $c_\beta=\cos\beta$, $u\equiv
\tan\beta=v_2/v_1$,
$h_i=1/\sqrt{2}\,(v_i+\tilde h_i)$, $m_Z^2=(g_1^2+g_2^2)\,v^2/4$.
At large $\tan\beta$:
\medskip
\bea\label{bb5}
 -v^2=\frac{2\tilde
  m_2^2}{\lambda_2}+\frac{1}{u\,\lambda_2^2}\,(-4\,m_3^2\lambda_2-
8\,\lambda_7\,\tilde m_2^2)+\cO(1/u^2);
 \eea

\bigskip
\noindent
Definition (\ref{delta2}) obtained using (\ref{min})
can be used to find  the most general result
$\Delta_p$  for the EW fine-tuning wrt a parameter $p$. 
This takes account of the dependence $\beta=\beta(p)$
induced by the min conditions. One finds
the  general expression:
\bigskip
\bea\label{ggg}
\Delta_p=\frac{\partial \ln v^2}{\partial\ln p}=
\frac{-\,\,\big\{ 2 \,w_1'\,z_1-(1/4)\,z_1'\, w_2
+[ w_3'+ (1/v^2)\,z_2']\,\,[z_3
+v^2 \,w_4]\,\big\}}{
-(1/32)\,v^2 \,w_2^2 - w_3\,[- z_3 -w_4\,v^2]}
\eea

\bigskip\noindent
with the following notations:

\bea
w_1'&\equiv
&\lambda_6'\cos^4\beta+\lambda_{3451}'\,\cos^3\beta\sin\beta
-\frac{3}{4}\,(\lambda_6'-\lambda_7')\,\sin^2
2\beta-\lambda_{3452}'\,\cos\beta
\sin^3\beta-\lambda_7'\,\sin^4\beta,
\nonumber\\[4pt]
w_2&\equiv \!\!
& 4\,(\lambda_6+\!\lambda_7)\cos 2\beta\!+\!
4\,(\lambda_6-\lambda_7)\,\cos
4\beta- 2 [\lambda_1-\lambda_2+(\lambda_1+\lambda_2-2\,\lambda_{345})
\cos 2\beta]\,\sin 2\beta
\nonumber\\[4pt]
w_3&\equiv &
\frac{1}{2}\,\lambda_1\cos^4\beta+2\lambda_6\,\cos^3\beta\sin\beta+
\frac{1}{4}\lambda_{345}\sin^2
2\beta+2\lambda_7\cos\beta\sin^3\beta+\frac{1}{2}
\lambda_2\sin^4\beta\qquad\,\,\,
\nonumber\\[4pt]
w_3'&\equiv 
&\frac{1}{2}\,\lambda_1'\cos^4\beta+2\lambda_6'\,\cos^3\beta\sin\beta+
\frac{1}{4}\lambda_{345}'\sin^2
2\beta+2\lambda_7'\cos\beta\sin^3\beta+\frac{1}{2}
\lambda_2'\sin^4\beta
\nonumber\\[4pt]
w_4&\equiv &
-(\lambda_1+\lambda_2-2\lambda_{345})\,\cos 4\beta-2
\,(\lambda_6+\lambda_7)\,\sin 2\beta+4 \,(\lambda_7-\lambda_6)\,\sin 4\beta
\nonumber\\[4pt]
z_1&\equiv & -2 m_3^2\cos 2\beta +(\tilde m_2^2-\tilde m_1^2)\,\sin 2\beta,
 \nonumber\\[4pt]
z_1'&\equiv&  -2 (m_3^2)'\cos 2\beta +[(\tilde m_2^2)'-(\tilde
  m_1^2)']\,\sin 2\beta
\nonumber\\[4pt]
z_2'&\equiv& (\tilde m_1^2)' \cos^2\beta+(\tilde m_2^2)'\sin^2 \beta
-(m_3^2)'\,\sin 2\beta,
 \nonumber\\[4pt]
z_3&\equiv & \,\big[
4\,\tilde m_2^2 -4\,\tilde m_1^2+(\lambda_2-\lambda_1)\,v^2\big]
\,\cos 2\beta+8\,m_3^2\,\sin 2\beta\label{genD}
\eea

\medskip\noindent
where 
$\lambda_{345}=\lambda_3+\lambda_4+\lambda_5$;
$\,\,\, \lambda_{345j}\equiv
\lambda_3+\lambda_4+\lambda_5-\lambda_j$,\,\,$(j=1,2)$ and with:
\medskip
\bea
(m_3^2)'\equiv \frac{\partial m_3^2}{\partial \ln p},
\quad
(\tilde m_j^2)'\equiv \frac{\partial \tilde m_j^2}{\partial \ln p},
\,\,\,(j=1,2);
\quad
\lambda_i'\equiv \frac{\partial \lambda_i}{\partial\ln p},
\,\,\,i=1,2...7.
\eea

\medskip\noindent
The general result (\ref{ggg}), (\ref{genD}) can be applied to any 
two-Higgs doublet model, which includes all radiative corrections
in the couplings and soft masses. The result in
(\ref{ggg}) simplifies considerably
in most cases, since usually many $\lambda_i$ are independent of $p$, 
i.e. have $\lambda_i'=0$.

It is worth taking some particular limits of the above result for $\Delta_p$.
At large $\tan\beta$:
\medskip
\bea
\Delta_p&=& \frac{1}{v^2}\,\,\bigg\{
\frac{
v^2\,[\, 4 \,\lambda_7\,(m_3^2)'
- 4\lambda_7' \,m_3^2+\lambda_2'\,(\,2 \,\tilde m_1^2-2 \,\tilde m_2^2+
\lambda_{3452}\,v^2\,)]}{
-2\,\lambda_2\,(\tilde m_1^2-\tilde m_2^2)+
[-\lambda_2 \lambda_{3452}
+ 2\lambda_7^2\,]
\,v^2 }\nonumber\\[10pt]
&&\qquad\qquad +\,\,
\frac{ 2\,  [\, 2\, (\tilde m_1^2- \tilde m_2^2)+
\lambda_{3452}\,\,v^2]\,\,(\tilde m_2^2)'}{
-2\,\lambda_2\,(\tilde m_1^2-\tilde m_2^2)+
[-\lambda_2 \lambda_{3452}
+ 2\lambda_7^2\,]
\,v^2 }\bigg\}+\cO(1/\tan\beta)
\eea

\medskip\noindent
which for $\lambda_2'=\lambda_7'=0$ gives
\medskip
\bea\label{b10}
\Delta_p=\frac{-2\,}{\lambda_2\,v^2}\,\,
\bigg\{
(\tilde m_2^2)'-\frac{2\lambda_7\,v^2\,\,
\big[\,\lambda_2\,(m_3^2)'+\lambda_7 \,(\tilde m_2^2)'\big]}{
2\lambda_7^2\,v^2-\lambda_2\,[\lambda_{3452}\,v^2+2\,(\tilde
m_1^2-\tilde m_2^2)]}\bigg\}+\cO(1/\tan\beta)
\eea
In MSSM $\lambda_7=\lambda_7'=0$ then 
\begin{eqnarray}  \label{b11} 
\Delta_p=\frac{-2}{\lambda_2\,v^2}\,\, 
(\tilde  m_2^2)^{\prime }+\mathcal{O}(1/\tan\beta) 
\end{eqnarray} 

\medskip\noindent 
which is consistent with (\ref{bb5}) and the
definition of $\Delta_p$ (assuming $\delta'=0$).

\noindent
In MSSM with $d=5$ operators, $\lambda_7=\zeta_1$ so
\medskip
\bea\label{t4}
\Delta_{p}=-\frac{2}{v^2\,(1+\delta)\, m_Z^2}
\bigg[
(\tilde  m_2^2)'
-\frac{\zeta_1\,v^2}{\tilde m_2^2-\tilde m_1^2+m_Z^2 \,(1+\delta/2)}
(m_3^2)'\bigg]
+\cO(1/\tan\beta)
\eea

\medskip
\noindent
which recovers the results of (\ref{five}).


\begin{thebibliography}{99} 
\bibitem{Dimopoulos:1981yj} 
S.~Dimopoulos, S.~Raby and F.~Wilczek, 
``Supersymmetry And The Scale Of Unification,'' Phys.\ Rev.\ D \textbf{24} 
(1981) 1681. 
 
\bibitem{Dimopoulos:1981zb} 
S.~Dimopoulos and H.~Georgi, ``Softly Broken 
Supersymmetry And SU(5),'' Nucl.\ Phys.\ B \textbf{193} (1981) 150.  
 

 \bibitem{Ibanez:1981yh} 
L.~E.~Ibanez and G.~G.~Ross, ``Low-Energy 
Predictions In Supersymmetric Grand Unified Theories,'' Phys.\ Lett.\ B  
\textbf{105} (1981) 439. 
 
 \bibitem{Ghilencea:2001qq} 
D.~M.~Ghilencea and G.~G.~Ross, ``Precision 
prediction of gauge couplings and the profile of a string theory,'' Nucl.\ 
Phys.\ B \textbf{606} (2001) 101 [arXiv:hep-ph/0102306].  
 
\bibitem{Ibanez:1982fr} 
L.~E.~Ibanez and G.~G.~Ross, ``SU(2)-L X U(1) 
Symmetry Breaking As A Radiative Effect Of Supersymmetry Breaking In Guts,'' 
Phys.\ Lett.\ B \textbf{110} (1982) 215. 
``Supersymmetric Higgs and radiative electroweak breaking,'' Comptes Rendus 
Physique \textbf{8} (2007) 1013 
 
\bibitem{higgsboundLEP} 
R.~Barate \textit{et al.} [LEP Working Group for 
Higgs boson searches], ``Search for the standard model Higgs boson at LEP,'' 
Phys.\ Lett.\ B \textbf{565}, 61 (2003) [arXiv:hep-ex/0306033]; S.~Schael  
\textit{et al.} [ALEPH Collaboration], ``Search for neutral MSSM Higgs 
bosons at LEP,'' Eur.\ Phys.\ J.\ C \textbf{47}, 547 (2006) 
[arXiv:hep-ex/0602042]. 
 
\bibitem{r3} 
R.~Barbieri and G.~F.~Giudice, ``Upper Bounds On Supersymmetric Particle 
Masses,'' Nucl.\ Phys.\ B \textbf{306} (1988) 63;  
 
  
\bibitem{Barbieri:1998uv} R.~Barbieri and A.~Strumia, ``About the 
fine-tuning price of LEP,'' Phys.\ Lett.\ B \textbf{433} (1998) 63 
[arXiv:hep-ph/9801353]. 
 
 
\bibitem{Chankowski:1998xv} P.~H.~Chankowski, J.~R.~Ellis, M.~Olechowski and 
S.~Pokorski, ``Haggling over the fine-tuning price of LEP,'' Nucl.\ Phys.\ B  
\textbf{544} (1999) 39 [arXiv:hep-ph/9808275].  
 
\bibitem{Chankowski:1997zh} P.~H.~Chankowski, J.~R.~Ellis and S.~Pokorski, 
``The fine-tuning price of LEP,'' Phys.\ Lett.\ B \textbf{423} (1998) 327 
[arXiv:hep-ph/9712234]. 
 
\bibitem{Kane:1998im} G.~L.~Kane and S.~F.~King, ``Naturalness implications 
of LEP results,'' Phys.\ Lett.\ B \textbf{451} (1999) 113 
[arXiv:hep-ph/9810374]. 
 
\bibitem{Batra:2003nj} P.~Batra, A.~Delgado, D.~E.~Kaplan and T.~M.~P.~Tait, 
``The Higgs mass bound in gauge extensions of the minimal supersymmetric 
standard model,'' JHEP \textbf{0402} (2004) 043 [arXiv:hep-ph/0309149].  
 
\bibitem{Giudice:2006sn} G.~F.~Giudice and R.~Rattazzi, ``Living dangerously 
with low-energy supersymmetry,'' Nucl.\ Phys.\ B \textbf{757} (2006) 19 
[arXiv:hep-ph/0606105]. 
 
 \bibitem{Casas:2003jx} J.~A.~Casas, J.~R.~Espinosa and I.~Hidalgo, ``The 
MSSM fine tuning problem: A way out,'' JHEP \textbf{0401} (2004) 008 
[arXiv:hep-ph/0310137];
``A relief to the supersymmetric fine tuning problem,''
  arXiv:hep-ph/0402017.

 \bibitem{Dermisek:2007yt} R.~Dermisek and J.~F.~Gunion, ``The NMSSM Solution 
to the Fine-Tuning Problem, Precision Electroweak Constraints and the 
Largest LEP Higgs Event Excess,'' Phys.\ Rev.\ D \textbf{76} (2007) 095006 
[arXiv:0705.4387 [hep-ph]]. 
 
 
 \bibitem{Piriz:1997id} D.~Piriz and J.~Wudka, ``Effective operators in 
supersymmetry,'' Phys.\ Rev.\ D \textbf{56} (1997) 4170 
[arXiv:hep-ph/9707314]. 

 
\bibitem{Polonsky:2000zt} N.~Polonsky and S.~f.~Su, ``Low-energy limits of 
theories with two supersymmetries,'' Phys.\ Rev.\ D \textbf{63} (2001) 
035007 [arXiv:hep-ph/0006174]. 
 
\bibitem{M0} S.~P.~Martin, ``Dimensionless supersymmetry breaking couplings, 
flat directions, and the origin of intermediate mass scales,'' Phys.\ Rev.\ 
D \textbf{61} (2000) 035004 [arXiv:hep-ph/9907550],  

 
\bibitem{Barbieri:1999tm} R.~Barbieri and A.~Strumia, ``What is the limit on 
the Higgs mass?,'' Phys.\ Lett.\ B \textbf{462} (1999) 144 
[arXiv:hep-ph/9905281]. 
 
 
\bibitem{Dine} M.~Dine, N.~Seiberg and S.~Thomas, ``Higgs Physics as a 
Window Beyond the MSSM (BMSSM),'' Phys.\ Rev.\ D \textbf{76} (2007) 095004 
[arXiv:0707.0005 [hep-ph]]. 
 
 \bibitem{Antoniadis:2008es} I.~Antoniadis, E.~Dudas, D.~M.~Ghilencea and 
P.~Tziveloglou, ``MSSM with Dimension-five Operators (MSSM$_5$),'' Nucl.\ 
Phys.\ B \textbf{808} (2009) 155 [arXiv:0806.3778 [hep-ph]].  

\bibitem{Antoniadis:2009rn}
  I.~Antoniadis, E.~Dudas, D.~M.~Ghilencea and P.~Tziveloglou,
  ``MSSM Higgs with dimension-six operators,''
  arXiv:0910.1100 [hep-ph].

\bibitem{Blum:2008ym} K.~Blum and Y.~Nir, ``Beyond MSSM Baryogenesis,'' 
Phys.\ Rev.\ D \textbf{78} (2008) 035005 [arXiv:0805.0097 [hep-ph]].  
  
\bibitem{Carena:1995bx} M.~S.~Carena, J.~R.~Espinosa, M.~Quiros and 
C.~E.~M.~Wagner, ``Analytical expressions for radiatively corrected Higgs 
masses and couplings in the MSSM,'' Phys.\ Lett.\ B \textbf{355} (1995) 209 
[arXiv:hep-ph/9504316]. 
 
 
\bibitem{deBoer} W.~de Boer, R.~Ehret and D.~I.~Kazakov, ``Predictions of 
SUSY masses in the minimal supersymmetric grand unified theory,'' Z.\ Phys.\ 
C \textbf{67} (1995) 647 [arXiv:hep-ph/9405342].  
 
\bibitem{r1} L.~E.~Ibanez, C.~Lopez and C.~Munoz, ``The Low-Energy 
Supersymmetric Spectrum According To N=1 Supergravity Guts,'' Nucl.\ Phys.\ 
B \textbf{256} (1985) 218; 
 
\bibitem{r2}
L.~E.~Ibanez and C.~Lopez, ``N=1 Supergravity, The Weak Scale And The 
Low-Energy Particle Spectrum,'' Nucl.\ Phys.\ B \textbf{233} (1984) 511;  

\bibitem{r4} 
L.~E.~Ibanez and C.~Lopez, ``N=1 Supergravity, The Breaking Of SU(2) X U(1) 
And The Top Quark Mass,'' Phys.\ Lett.\ B \textbf{126} (1983) 54.  
 
\bibitem{TeV} The Tevatron Electroweak Working Group (TevEWWG) and CDF 
Collaboration and D0 Collab, ``A Combination of CDF and D0 Results on the 
Mass of the Top Quark,'' arXiv:0803.1683 [hep-ex].  
 
 \bibitem{Espinosa:1998re} J.~R.~Espinosa and M.~Quiros, ``Gauge unification 
and the supersymmetric light Higgs mass,'' Phys.\ Rev.\ Lett.\ \textbf{81} 
(1998) 516 [arXiv:hep-ph/9804235]. 
 
 
\bibitem{Espinosa:1991gr} J.~R.~Espinosa and M.~Quiros, ``On Higgs Boson 
Masses In Nonminimal Supersymmetric Standard Models,'' Phys.\ Lett.\ B  
\textbf{279} (1992) 92. 
 
\bibitem{Espinosa:1991wt} J.~R.~Espinosa and M.~Quiros, ``Higgs triplets in 
the supersymmetric standard model,'' Nucl.\ Phys.\ B \textbf{384} (1992) 
113. 
 
\bibitem{kaplan} A. Maloney, A. Pierce, J. Wacker, ``D-terms, unification, 
and the Higgs mass'' JHEP \textbf{06} (2006) 034. 
 
\bibitem{Bellazzini:2009ix} B.~Bellazzini, C.~Csaki, A.~Delgado and 
A.~Weiler, ``SUSY without the Little Hierarchy,'' arXiv:0902.0015 [hep-ph].  
\end{thebibliography}
\end{document}